\PassOptionsToPackage{sort&compress}{natbib}
\documentclass{article}
\usepackage{natbib}
\usepackage{geometry}
\usepackage{pgfplots} 
\usepackage{tikz}
\usepackage{xcolor}
\usepackage{helvet}
\pgfplotsset{compat=1.17}
\usepackage{graphicx, adjustbox}
\usepackage{subfig}
\usepackage{pifont}
\usepackage{float}
\usepackage{caption}
\usepackage{afterpage}
\usepackage{graphicx} 
\usepackage{amssymb}
\usepackage{amsmath}
\usepackage{capt-of}
\usepackage{multirow}
\usepackage[affil-it]{authblk}
\usepackage{hyperref}
\usepackage{lineno}
\definecolor{darkgreen}{rgb}{0.0, 0.5, 0.0}
\definecolor{verylightorange}{RGB}{255, 235, 204}

\DeclareUnicodeCharacter{2212}{-}

\begin{document}

\title{SemanticST: Spatially Informed Semantic Graph Learning for Clustering, Integration, and Scalable Analysis of Spatial Transcriptomics}
\author[1]{Roxana Zahedi}
\author[2,3]{Ahmadreza Argha}
\author[2,3]{Nona Farbehi}
\author[1]{Ivan Bakhshayeshi}
\author[4]{Youqiong Ye}
\author[2,3]{Nigel H. Lovell}
\author[1,3]{Hamid Alinejad-Rokny \thanks{Corresponding author: \href{ h.alinejad@unsw.edu.au}{ h.alinejad@unsw.edu.au}, Tel-Fax: +61 2 9385 3911}}
\affil[1]{UNSW BioMedical Machine Learning Lab (BML), School of Biomedical Engineering, UNSW Sydney 2052, NSW, Australia}
\affil[2]{School of Biomedical Engineering, UNSW Sydney, 2052, NSW, Australia}
\affil[3]{Tyree Institute of Health Engineering (IHealthE), UNSW Sydney, 2052, NSW,Australia}
\affil[4]{Center for Immune-Related Diseases at Shanghai Institute of Immunology, Shanghai Jiao Tong University School of Medicine, Shanghai, 200001 China}
\maketitle

\begin{abstract}
Spatial transcriptomics (ST) technologies enable gene expression profiling with spatial resolution, offering unprecedented insights into tissue organization and disease heterogeneity. However, current analysis methods often struggle with noisy data, limited scalability, and inadequate modelling of complex cellular relationships.
We present SemanticST, a biologically informed, graph-based deep learning framework that models diverse cellular contexts through multi-semantic graph construction. SemanticST builds multiple context-specific graphs—capturing spatial proximity, gene expression similarity, and tissue domain structure, and learns disentangled embeddings for each. These are fused using an attention-inspired strategy to yield a unified, biologically meaningful representation. A community-aware min-cut loss improves robustness over contrastive learning, particularly in sparse ST data.
SemanticST supports mini-batch training, making it the first graph neural network scalable to large-scale datasets such as Xenium (500,000+ cells). Benchmarking across four platforms (Visium, Slide-seq, Stereo-seq, Xenium) and multiple human and mouse tissues shows consistent 10–20\% gains in ARI, NMI, and trajectory fidelity over DeepST, GraphST, and IRIS.
In re-analysis of breast cancer Xenium data, SemanticST revealed rare and clinically significant niches, including triple receptor-positive clusters, spatially distinct DCIS-to-IDC transition zones, and FOXC2+ tumour-associated myoepithelial cells, suggesting non-canonical EMT programs with stem-like features. Without relying on multi-omic integration, SemanticST also achieved vertical and horizontal alignment across samples, correcting batch effects while preserving biological variation, and outperforming Seurat+Harmony in iLISI and cLISI metrics.
SemanticST thus provides a scalable, interpretable, and biologically grounded framework for spatial transcriptomics analysis, enabling robust discovery across tissue types and diseases, and paving the way for spatially resolved tissue atlases and next-generation precision medicine.
\end{abstract}

\section*{Introduction}
Exploring biomolecules and understanding their functions require advanced methods for measurement, followed by computational biology for analysis. While genome-wide methods such as single cell RNA sequencing (RNA-seq) measure a broad range of biomolecules across the entire genome, they often come with the limitation of losing spatial context within tissues due to the need to dissociate cells from their native environment, which can induce stress, cell death, and loss of important spatial information \cite{williams2022introduction}. Moreover, measuring transcriptomes while preserving spatial context is critical for characterizing transcriptional patterns and tissue regulation, as the localization of mRNA within a cell significantly affects how gene activity is regulated in different cellular regions \cite{williams2022introduction, mrNALOC}. Recent innovations in oligonucleotide synthesis and fluorescence microscopy have enabled the development of advanced technologies for the detailed mapping of gene expression and transcript localization within tissues at the single-cell level. These methods, collectively referred to as spatially resolved transcriptomics (SRT) or simply spatial transcriptomics (ST), are transforming our understanding of cellular organization. \\ Broadly, ST methods can be divided into two main categories. First, (i) image-based methods preserve spatial information by imaging mRNAs in situ. These can be further subdivided into: (A) hybridization-based approaches, such as in situ hybridization (ISH), which include seqFISH \cite{lubeck2014single}, seqFISH+ \cite{maynard2021transcriptome}, and MERFISH \cite{chenkh2015rnaimaging}; and (B) sequencing-based approaches, such as in situ sequencing (ISS), which include STARmap \cite{wang2018three}, where amplified mRNAs are sequenced in situ. While both methods provide high spatial resolution, even at the subcellular level, efficiency decreases as the number of profiled genes increases, primarily due to the extended imaging time required for high-throughput microscopy \cite{eng2019transcriptome}. Second, (ii) sequencing-based approaches that profile mRNA from intact tissue by next-generation sequencing techniques. Sequencing-based methods can be categorized as (A) array-based methods such as ST \cite{staahl2016visualization}, Slide-seq \cite{stickels2021highly}, and HDST \cite{vickovic2019high} , which directly capture spatial locations using grid-like surfaces containing attached probes that indicate mRNA positions, and (B) microdissection-based methods such as laser capture microdissection (LCM) \cite{espina2006laser}, which use techniques such as LCM or microfluidics to isolate specific regions or cells while preserving spatial information. The untargeted nature of sequencing-based methods, along with their capacity to profile larger tissue sections compared to image-based techniques, enables comprehensive transcriptomic profiling across diverse organismal tissue types, without requiring image preprocessing. However, majority of these methods offer lower spatial resolution than image-based approaches, as they provide gene expression data at the level of 'spots', each covering multiple cells, potentially masking finer details of cellular heterogeneity and tissue architecture.\\
Regardless of the specific ST technologies used, the data generated by these techniques are complex, vast, and multi-modal, as they integrate gene expression profiles, spatial information, and histological images. The rapid advancement of ST technologies necessitates the development of efficient machine learning and deep learning approaches to facilitate their analysis. Whether or not methods incorporate spatial information, recent approaches generally rely on techniques to obtain latent representations or embedding of the data, followed by downstream analysis on the resulting embeddings. 

Notably, the concept of embedding refers to any transformation of high-dimensional data into a lower-dimensional space, while preserving essential information and relationships between data points, ranging from simpler methods like principal component analysis (PCA) to more sophisticated approaches, such as deep graph neural networks (GNNs). To obtain embeddings for ST data, various methodologies have been deployed in recent approaches. Besides Seurat \cite{satija2015spatial}, which does not incorporate spatial data, most recent approaches utilize spatial information in their analyses and can be categorized into two classes: (i) model architecture enhancement approaches, which modify methodological aspects. For example, Giotto \cite{dries2021giotto} employs a hidden Markov random field (HMRF) model to identify coherent spatial patterns from the neighbouring regions of each gene expression profile. SpaGCN \cite{hu2021spagcn} utilizes a graph convolutional neural network (GCN), while DeepST \cite{xu2022deepst} employs a denoising autoencoder combined with a domain adversarial neural network. STAGATE \cite{STAGATE} uses a deep graph attention mechanism along with a cell-type-aware module, and SEDR \cite{SEDR} integrates a deep autoencoder model with a variational graph autoencoder network to obtain embeddings. (ii) Learning paradigm shifts, which introduce changes to the learning strategy. Methods in this class do not include learning strategies that arise from model differences, such as Kullback-Leibler (KL) divergence loss, which is mostly used when approaches employ variational graph autoencoders or compare distributions. Instead, they focus on providing new learning strategies that impact embeddings, such as the development of novel loss functions or enhancements to input data. For example, SpaGCN adds a new dimension to the two-dimensional spatial coordinates by incorporating the RGB channels from histology images, while stLearn \cite{pham2023robust} computes a pseudo-time-space distance that integrates gene expression and spatial distance between sub-clusters. GraphST \cite{graphST} and conST \cite{const} employ self-supervised contrastive learning methods with different strategies for handling negative samples. Recent approaches, such as Spatial-MGCN, have attempted to modify the input graph structure by using multiple similarity metrics between cells.\\ 

However, current existing methods have limitations in addressing various aspects of ST data analysis. The current limitations of the aforementioned methods can be assessed from three perspectives: (i) data representation, (ii) learning strategy, and (iii) scalability. From the data representation perspective, incorporating spatial information often leads to the use of graph structures as the primary model for data representation. However, these methods typically rely on a fixed neighbourhood size when constructing adjacency matrices, regardless of tissue type or species. This approach overlooks the complexity and non-uniformity of cellular relationships within tissues, where cells exhibit intricate patterns of interaction. These relationships vary based on context—cells may exhibit strong associations in one functional aspect but diverge significantly in another—emphasizing the need to better understand and capture these diverse interactions. 
From the learning strategy perspective, since current ST data are unlabeled, tasks such as clustering, cell-type deconvolution, and imputation are unsupervised. As a result, efficient unsupervised learning strategies, such as self-supervised learning, are needed to tackle these challenges. This has led to recent innovations like GraphST, which utilizes contrastive learning strategies with the infomax loss function \cite{velickovic2019deep}. However, the infomax loss constructs positive pairs from the input graph and negative pairs by corrupting it, resulting in longer training times due to the need to learn from both types of pairs. Moreover, the effectiveness of the results is highly dependent on the corruption strategy. Additionally, embeddings generated by these methods often require additional steps, such as clustering, to derive insights, rather than directly providing meaningful information.\\
Lastly, recent deep learning methods have emphasized spatial contributions by incorporating graph structures, where the entire graph representing spatial dependencies is input into the model to capture both local and global relationships. However, with advancements in ST, particularly with high-resolution platforms such as single-cell resolution, current approaches face challenges in effectively managing large ST datasets like those produced by the Xenium platform. Addressing this issue requires a scalable and computationally efficient approach that enables mini-batch training while preserving both local and global information across ST data. Although SGCAST \cite{li2024sgcast} introduced mini-batch training, it represents gene expression similarity by calculating the Euclidean distance between principal components derived from gene expression data. This is notable because, from a technical perspective, high-dimensional data like PCA-transformed data may not naturally conform to Euclidean geometry \cite{beyer1999nearest}.\\

Aside from smFISH- and ISS-based techniques, which are applicable to thick tissue sections, all other types of ST techniques are limited by section thickness, which is usually at least 10 $\mu m$ for frozen sections to allow proper permeabilization and RNA capture \cite{moses2022museum}. This limitation necessitates the analysis of vertical consecutive ST sections or 3D analysis to learn their joint representation and perform vertical integration. Another challenge arises from the spatial warping between different slices, which is inevitable due to measurement errors and requires slice alignment to transform the various slices into a shared spatial location \cite{liu2024computational}. For simplicity, we refer to this as horizontal integration between two adjacent ST slices. This concept differs from the horizontal integration between single-cell RNA sequencing (scRNA-seq) and ST, as explained by Argelaguet et al.\cite{argelaguet2021computational}. Recently developed methods are mostly limited to analyzing single tissue slices and cannot be effectively applied to multiple vertical and horizontal ST slices, where batch effect removal becomes a critical step for integrating serial tissue sections. Approaches like SEDR have attempted to use batch effect correction methods such as Harmony \cite{korsunsky2019fast} and scVI \cite{lopez2018deep} on their latent embeddings. However, these tools were originally designed for scRNA-seq, relying solely on gene expression, and are not well-suited for ST data. While STAGATE introduces a method to account for spatial information across slices, its performance may be compromised when batch effects are pronounced and require explicit correction. Among the available methods, DeepST and GraphST are the only ST-specific approaches that incorporate batch effect correction in their embeddings. However, their performance is also hindered by the aforementioned limitations related to data representation and learning strategies, as discussed earlier.\\

To address these limitations, We propose \textbf{SemanticST}, a novel unsupervised learning approach based on community detection and semantic graph structure learning, designed to tackle the complexities of modelling biological processes, including clustering and integration, in spatial ST data.
SemanticST uses a learnable weighted graph representation, termed semantic graphs, which enhances the model's ability to integrate and analyse complex biological data in a spatial context. In this approach, we employ a fixed number of semantic graphs, each designed to capture different aspects of the input data. For each semantic graph, a unique embedding is learned using an autoencoder with graph convolutional network (GCN) layers, representing distinct semantic features in latent space. To combine these representations, we introduce a learnable weight, referred to as the semantic score, for each semantic graph. The final graph representation is then dynamically fused by weighting and combining the individual embeddings based on their semantic scores, resulting in a more accurate and adaptive graph representation.
Furthermore, drawing inspiration from the community detection problem \cite{DUONG2023109126}, we incorporate the mincut loss function together with the Deep MinCut (DMC) learning algorithm. This approach not only captures the global structure of the graph but also reduces redundant training time. More importantly, it ensures that the learned embeddings are both meaningful and interpretable, providing a more robust and insightful representation of the graph. 
Notably, we incorporated a mini-batch training option in SemanticST by training the model on spatial graphs in smaller batches, allowing the learned semantics graph to maintain both local and global perspectives across batches. This feature makes SemanticST memory-efficient and scalable, enabling its application to any ST dataset, regardless of the number of samples.
We assessed SemanticST across various ST datasets with varying resolutions, including 10X Visium (human and mouse brain, as well as breast cancer), Slide-seqV2 (mouse hippocampus), Stereo-seq (mouse olfactory bulb and mouse embryo), and Xenium (breast cancer). The clustering results of SemanticST, supported by the heatmap of learned semantic scores, demonstrate its remarkable capability to identify spatial domains, outperforming recent state-of-the-art methods. SemanticST not only excels in delineating spatial domains but also captures the biological significance embedded within each semantic graph. Moreover, the latent representation obtained from multiple vertical tissue slices of the human brain and mouse breast cancer, alongside horizontal slices of the mouse brain, shows that SemanticST effectively utilizes spatial information to identify spatial domains across tissues, while addressing batch effects comprehensively, without relying on predefined batch factors.
\\

\section*{Results}
\subsection{Overview of SemanticST}
\begin{figure*}[h!]
    \centering
     \begin{adjustbox}{max width=\textwidth}
           \begin{tikzpicture}
        
        \node at (10,10) {\includegraphics[scale=1]{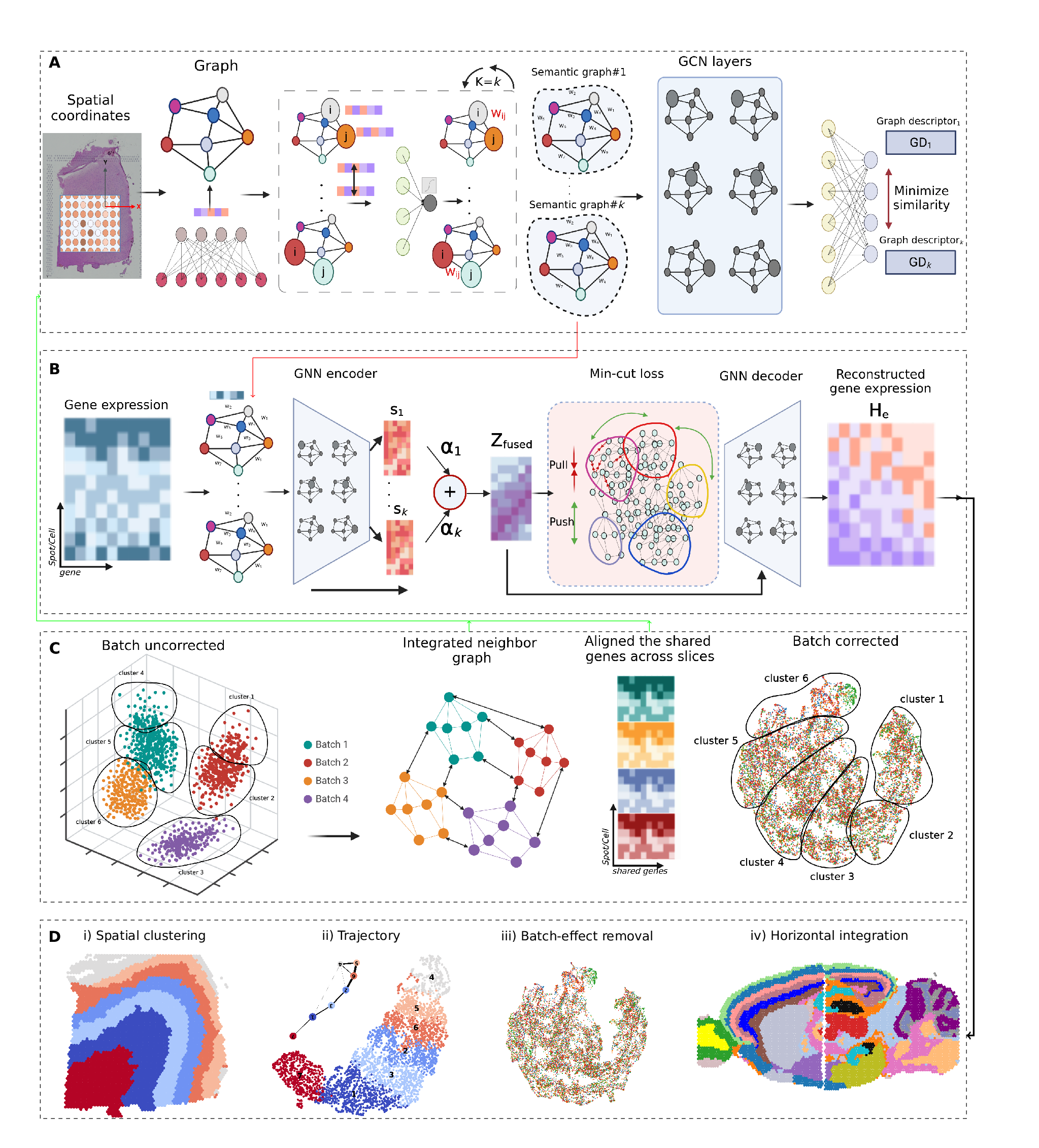}};

        \end{tikzpicture}
    \end{adjustbox}
   \phantomcaption
    \label{fig:semantic}
\end{figure*}

SemanticST comprises two main steps: semantic graph learning (Fig.~\ref{fig:semantic}A) and learning embedding space (Fig.~\ref{fig:semantic}B). In the first step, SemanticST uses spatial information to construct a neighbourhood graph, which represents the spatial dependency of spots or cells. Each node in the graph is assigned a feature corresponding to the gene expression vector of the respective cell or spot. This step is important to extract multiple graphs, treating them as semantic graphs to ensure that each graph represents a different aspect of the data.
In this step, SemanticST emphasises the algorithm’s ability to capture various aspects of ST data, mimicking the complex relationships between cells across the tissue. As a result, at the end of this step, we obtain multiple semantic graphs with shared structures and gene expression but differing graph weights.
In the second step, SemanticST utilises the learned semantic graphs and the gene expression matrix to learn the latent representation of the input ST data. The input in this step consists of $K$ weighted graphs with the same adjacency matrix and a shared gene expression matrix for all semantic graphs. By passing these semantic graphs through an encoder, SemanticST learns $K$ embedding vectors that serve as latent representations for each semantic graph. SemanticST then fuses these embeddings by assigning a learnable score to each one, combining them into a final latent representation, which is a weighted sum of all embeddings.
Given that adjacent spots or cells typically share similar patterns, SemanticST applies an unsupervised approach using a community-based loss function, as an alternative to contrastive learning. This approach generates several communities, each representing a local context of the spatial data, and assigns embeddings to the appropriate community. This ensures that the local context is preserved by enforcing similar embeddings for adjacent spots or cells.
Finally, SemanticST passes the fused embeddings to a decoder to reconstruct the input gene expression matrix. To achieve this, SemanticST utilises a reconstruction loss function by feeding the decoder's output, $H_e$, back into the network and calculating the reconstruction loss to preserve biological information. During the optimisation of the network in Step B, SemanticST jointly minimises the self-reconstruction loss and the min-cut loss. Note that the first and second steps (A and B,respectively) are independently executed.\\
In vertical integration (Fig.~\ref{fig:semantic}C), the main challenge is removing batch effects between consecutive slices, which arise due to tissue thickness and technological limitations in ST data. As shown in Fig.~\ref{fig:semantic}C, SemanticST constructs an integrated neighbourhood graph by combining spatial information from all slices and concatenates the gene expression matrices, preserving shared genes across slices. The final graph and concatenated gene expression matrix are then passes to Step A, after which the remaining processes are similar to single-slice learning. SemanticST is able to implicitly remove batch effects without the need for predefined batch factors. It not only learns latent representations for each slice but also captures joint representations across slices by leveraging both spatial information and gene expression data. For horizontal integration, two slices are stitched together horizontally, and the process follows the same workflow as single-slice analysis.
The final output of SemanticST after training is $H_e$, which can be utilised for downstream analyses, including spatial clustering, trajectory inference, and data integration (Fig.~\ref{fig:semantic}D).
\setcounter{figure}{1}
\afterpage{ \clearpage
    \noindent\rule{\textwidth}{0.4pt} 
    \captionsetup[figure]{labelformat=empty} 
    \begin{minipage}{\textwidth}
     \captionof{figure}{\textbf{Fig. 1|} \textbf{Overview of SemanticST.}  \textbf{A.} SemanticST uses the spatial coordinates of captured transcriptomics to construct a neighbour graph, where each node represents a cell or spot, and the edges represent spatial dependencies between them. The feature of each node is a vector of genes expressed in that spot or cell. Using this graph, SemanticST generates $k$ semantic graphs to learn different semantic representations, with each graph capturing a specific aspect of the data. The algorithm is designed to maximize the distance between these semantic graphs, ensuring they encode distinct information. \textbf{B.} Using the learned semantic graphs from Step A and the original gene expression matrix, SemanticST employs $k$ encoders with GCN layers to learn latent representations corresponding to each semantic graph. A learnable semantic score is then assigned to each latent representation to compute the fused embedding, $Z_{fused}$, by integrating information from all semantic graphs. Subsequently, an asymmetric decoder is used to reconstruct the gene expression matrix, producing the decoder output, $H_e$. SemanticST identifies communities in the embedding space and optimises the latent representations to align with the appropriate communities by minimising intra-community distances and maximising inter-community distances.
    \textbf{C.} This workflow represents batch correction by SemanticST. In this step, the shared neighbour graph is constructed, integrating spatial alignment and the shared gene expression matrix, which are then passed to SemanticST. The output of SemanticST is batch-corrected integrated data, which can subsequently be used for clustering and other downstream analyses, similar to a single slice. \textbf{D.} The decoder's output, $H_e$, is the final output of SemanticST, which can be utilised for downstream analyses, including spatial clustering, trajectory inference, batch effect removal, and horizontal integration.}
    \end{minipage}
    \noindent\rule{\textwidth}{0.4pt} 
    \vspace{1em}
    \addtocounter{figure}{-1} 
}

\subsection{Assessment of human dorsolateral prefrontal cortex (DLPFC) domain: SemanticST improve the identification of cortex layers}
\begin{figure*}[ht!]
    \centering
     \begin{adjustbox}{max width=\textwidth}
           \begin{tikzpicture}

            \node at (10,10) {\includegraphics[scale=1]{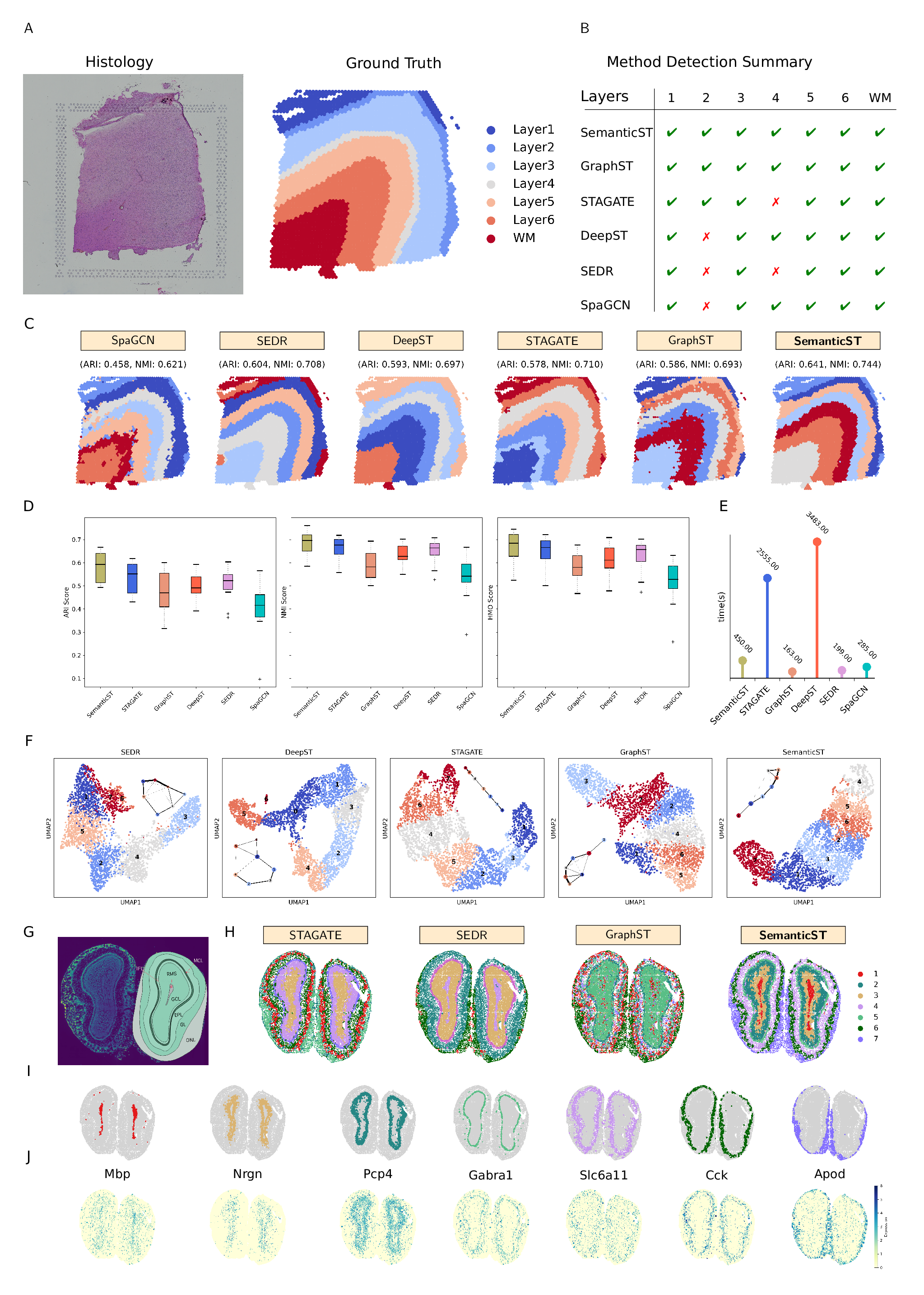}};
        \end{tikzpicture}
    \end{adjustbox}
    \phantomcaption
     \label{figure_DLPFC}
\end{figure*}

In our first evaluation, we applied SemanticST to a dataset from the 10x Visium platform comprising 12 sections from the human dorsolateral prefrontal cortex. This dataset, annotated by Maynard et al. \cite{maynard2021transcriptome}, is divided into five to seven cortical layers, as detailed in Supplementary Table 1, which served as the ground truth for our analysis. We assessed SemanticST's performance against state-of-the-art methods, including DeepST \cite{xu2022deepst}, SpaGCN \cite{hu2021spagcn}, SEDR \cite{SEDR}, GraphST \cite{graphST}, and STAGATE \cite{STAGATE}. Comparative metrics included adjusted Rand index (ARI), normalized mutual information (NMI), and homogeneity score (HOM). To ensure fair evaluation, clustering results were derived directly from embeddings without any post-processing, such as that applied by GraphST.
We first examined slice 151673 from the DLPFC dataset (Fig.~\ref{figure_DLPFC}). SemanticST demonstrated superior ARI and NMI scores while accurately recovering all cortical layers. Beyond metrics, we evaluated each method's ability to identify all brain layers. High ARI scores can occasionally mask issues where one layer is well-recovered but others are misclassified (Fig.~\ref{figure_DLPFC}B). For instance, SEDR achieved slightly lower ARI scores but failed to recover layer 2, merging it with layer 1. Similarly, DeepST showed reasonable ARI and NMI scores but did not detect layer 2 and misclassified portions of the white matter (WM) layer as separate clusters. STAGATE divided the WM into two layers and failed to identify layer 4. Although GraphST identified all clusters, it produced fragmented boundaries between them. In contrast, SemanticST not only achieved the highest ARI and NMI scores but also provided clearly defined and biologically accurate cluster boundaries. SpaGCN performed the poorest, with lower clustering scores and failure to recover layer 2.
Applying SemanticST across all 12 slices revealed its consistent superiority (Fig.~\ref{figure_DLPFC}D and Supplementary Fig.~S1). SemanticST achieved the highest median scores for ARI, NMI, and HOM across the dataset, outperforming all other methods. Moreover, SemanticST exhibited a narrower interquartile range (IQR) for these metrics, highlighting its robustness and consistency. In contrast, methods like SEDR and SpaGCN showed higher variability and significant outliers, particularly in ARI and HOM scores, suggesting inconsistent performance across slices. While STAGATE, GraphST, and DeepST demonstrated competitive median scores, none matched SemanticST in overall metrics, underscoring its superior performance.
Despite constructing four semantic graphs by aggregating information in the embedding space (Fig.~\ref{figure_DLPFC}E), SemanticST maintained a lower training time compared to STAGATE and DeepST, aligning with the efficiency of GraphST, SEDR, and SpaGCN. 
The DLPFC dataset provides a spatial map of gene expression reflecting the laminar organization of the human cerebral cortex, where each layer exhibits distinct gene expression patterns, morphology, and physiological \cite{klumpp2010review}. Fig.~\ref{figure_DLPFC}F illustrates UMAP embeddings generated by various methods alongside trajectory inference using the PAGA algorithm. SemanticST embeddings delineated well-separated cortical layers, and the corresponding trajectory preserved spatial relationships and chronological order from layer 1 to layer 6. This emphasizes SemanticST's effectiveness in capturing spatial and temporal patterns. By contrast, PAGA graphs for GraphST, DeepST, and SEDR failed to represent a linear cortical layer order. Although STAGATE produced the most linear trajectory among competing methods, it erroneously split the WM layer into two clusters.
To further investigate, we generated heatmaps of the learned attention scores for each semantic graph's embeddings, revealing the distinct features captured by each graph (Supplementary Figure S4A). Semantic Graph \#2 achieved high attention scores in layers 5 and 2. Notably, L2- and L5-enriched genes are strongly associated with schizophrenia (SCZD) \cite{maynard2021transcriptome}. Additionally, genes with de novo mutations linked to Autism Spectrum Disorder (ASD) \cite{satterstrom2020large} predominantly exhibit expression in L2 and L5 subsets, which are also enriched for genes implicated in common SCZD variants \cite{pardinas2018common} and, to a lesser extent, bipolar disorder (BPD) \cite{stahl2019genome}. These subsets, associated with specific clinical characteristics,  and can be further divided into distinct layers within the brain \cite{maynard2021transcriptome}. Therefore, our method not only generates highly informative embeddings for clustering and downstream analysis but also effectively captures biological signals, including those related to psychological disorders.
\afterpage{
    \clearpage
    
    \noindent\rule{\textwidth}{0.4pt} 
    \captionsetup[figure]{labelformat=empty} 
    \begin{minipage}{\textwidth}
    \captionof{figure}[Example Figure]{\textbf{Fig.2 |} \textbf{A} Histology image and corresponding ground truth from the original study of slice 151673. \textbf{B} Visual comparison of SemanticST and state-of-the-art methods (GraphST, STAGATE, DeepST, SEDR, and SpaGCN) in detecting each cortex layer in the DLPFC dataset. \textbf{C} Clustering results of these methods on slice 151673, along with their respective ARI and NMI scores. \textbf{D} Box plot displaying clustering metrics, including ARI, NMI, and HMO, for SemanticST and other methods across all DLPFC slices. \textbf{E} Comparison of running times for different methods on all DLPFC slices. \textbf{F} UMAP and PAGA plots of each method's embedding. \textbf{G} Tissue annotation derived from the study by Fu et al., showing corresponding laminar layers including RMS, GCL, IPL, MCL, EPL, and ONL. \textbf{H} Clustering results from STAGATE, SEDR, GraphST, and SemanticST on the mouse olfactory bulb dataset. \textbf{I} Detailed visual comparison of spatial domains identified by SemanticST and other methods. \textbf{J} Well-established marker genes supporting each cluster. }
    \end{minipage}
    \noindent\rule{\textwidth}{0.4pt} 
    \vspace{1em}
    \addtocounter{figure}{-1} 
}
\subsection{SemanticST more accurately delineated the laminar structure of the mouse olfactory bulb in the Stereo-seq dataset}
In the following analysis, we evaluated SemanticST on the mouse olfactory bulb tissue dataset from Stereo-seq, comparing the spatial domains identified by SemanticST to those obtained by GraphST, SEDR, and STAGATE. This dataset, annotated by Fu et al. \cite{SEDR}, defines seven distinct laminar layers in a DAPI-stained image, including the rostral migratory stream (RMS), granule cell layer (GCL), internal plexiform layer (IPL), mitral cell layer (MCL), external plexiform layer (EPL), and olfactory nerve layer (ONL) (Fig.~\ref{figure_DLPFC}G). Fig.~\ref{figure_DLPFC}H shows the clustering results of four methods, demonstrating that all methods successfully identified the outer layer domain (ONL) and approximately captured the laminar structure of the data. To explore the differences more closely, we separated each spatial domain corresponding to the laminar layers in Supplementary Figure S2C. GraphST produced scattered clusters for the EPL, combined the GCL and IPL into a single cluster, and failed to detect the RMS and GL layers. STAGATE identified all domains except for RMS, which it merged with GCL. While SemanticST accurately detected all laminar layers, including the outermost layers, SEDR successfully identified the RMS layer but could not distinguish the GCL and GL layers. This demonstrates that SemanticST (Fig.~\ref{figure_DLPFC}I) was the only method to accurately detect all spatial domains, with its clusters aligning well with known marker genes associated with each layer (Fig.~\ref{figure_DLPFC}J). Additionally, SemanticST effectively captured marker gene expression across regions, such as \textit{Mbp} and \textit{Slc6a11}, despite their overlap with other regions (Fig.~\ref{figure_DLPFC}J). This highlights the advantage of leveraging both spatial information and semantic scores. These findings confirm SemanticST's ability to delineate complex tissue structures with higher fidelity than competing methods, enabling deeper biological insights and advancing the analysis of ST data.

\subsection{SemanticST unveils spatial domains across scales: From adult mouse brain sections to hippocampal precision in Visium and Slide-seqV2 data }
\begin{figure*}[ht!]
    \centering
    \begin{adjustbox}{max width=\textwidth}
        \begin{tikzpicture}
   
             \node at (10,10) {\includegraphics[scale=1]{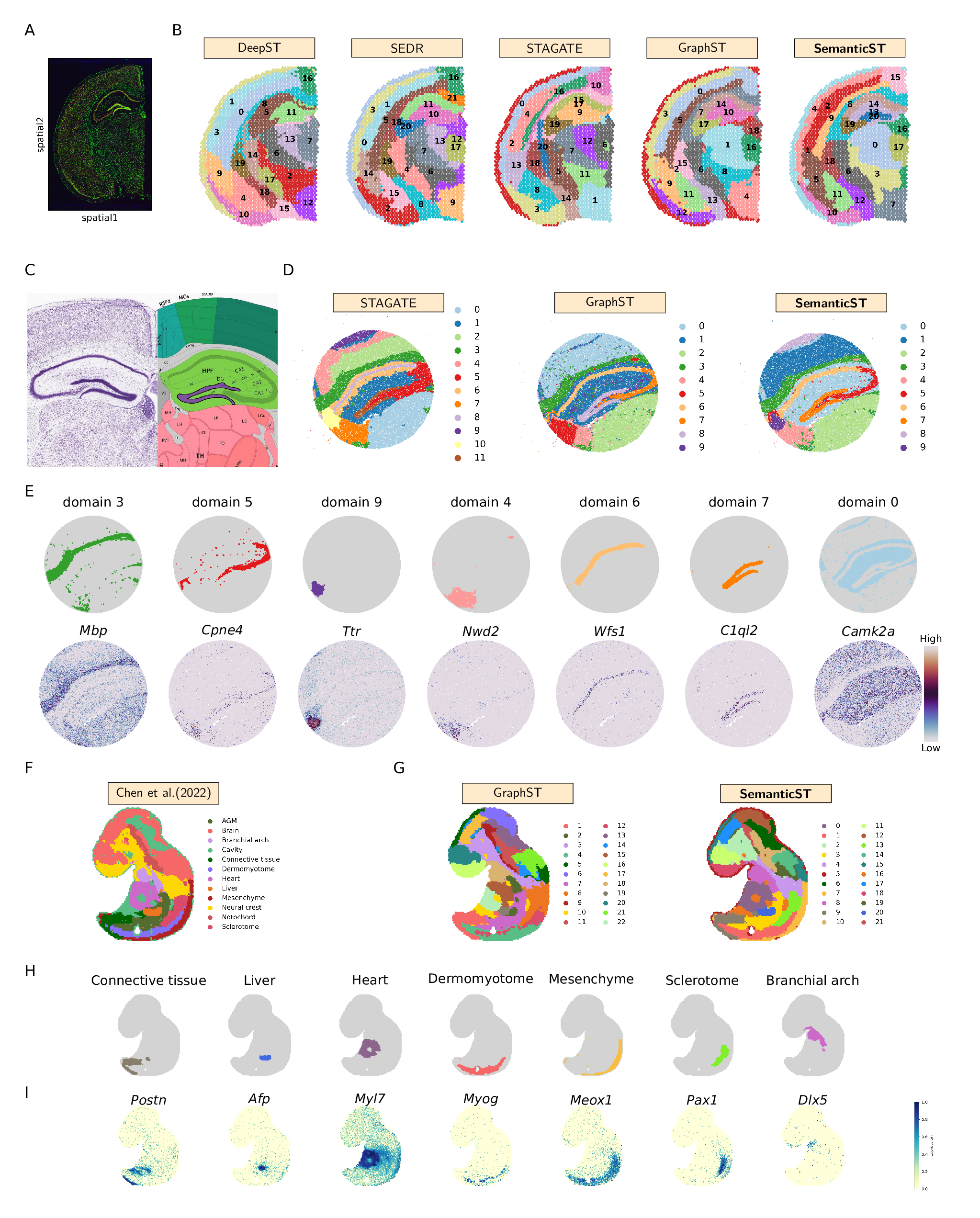}};

        \end{tikzpicture}
    \end{adjustbox}
     \phantomcaption
    \label{figure_mouse_brain}
\end{figure*}

For our subsequent analysis, we evaluated SemanticST's performance on more heterogeneous tissue encompassing multiple spatial domains. Our aim was to glean initial biological insights from each semantic graph and assess whether the learned semantic graph could effectively capture biological regions. To this end, we applied SemanticST to a 10x Visium dataset from a coronal mouse brain section (Fig.~\ref{figure_mouse_brain}.A) and compared its performance with other state-of-the-art methods including DeepST, SEDR, STAGATE, and GraphST (Fig.~\ref{figure_mouse_brain}.B).
Given the unknown number of clusters in this dataset, we employed the Louvain clustering algorithm \cite{blondel2008fast} on all embeddings obtained by the aforementioned algorithms. This approach ensured that any observed differences stemmed from the embedding space rather than clustering parameters. To facilitate comparison, we approximately categorized the domains identified by SemanticST into five regions based on the Allen Brain Atlas (Supplementary Figure S4B): cerebral cortex (domains 1, 2, 4, 5, 9, 10, and 11), fiber tracts (domains 6 and 8), hippocampal formation (domains 13, 14, 19, and 20), thalamus (domain 0), and hypothalamus (domains 3, 7, 16, and 17).
In the cerebral cortex region, SemanticST, GraphST, and SEDR demonstrated better identification of layer structure compared to DeepST and STAGATE. Regarding the hippocampal formation, which includes the curved, layered structure known as Ammon's horn (comprising CA1, CA2, and CA3) and the adjacent V-shaped formation called the dentate gyrus (DG) (including granule cell layer [DG-sg] and molecular layer [DG-mo]), all methods successfully detected these distinctive structures. However, DeepST failed to discern small spatial domains within Ammon's horn, instead detecting a single cluster for this area (domain 5). While SEDR, STAGATE, and GraphST identified subregions within Ammon's horn (CA1 and CA2), they could not distinguish between the layers of the DG (DG-sg and DG-mo). Notably, SemanticST was the only method to accurately identify CA1 (domain 19) and CA3 (domain 14) within Ammon's horn, as well as DG-sg (domain 20) and DG-mo (domain 13) regions within the dentate gyrus, demonstrating its superior performance in delineating small spatial domains within these complex anatomical structures. 

\afterpage{
    \clearpage
   
    \noindent\rule{\textwidth}{0.4pt} 
    \captionsetup[figure]{labelformat=empty} 
    \begin{minipage}{\textwidth}
    \captionof{figure}[Example Figure]{\textbf{Fig.3 |} \textbf{A} Immunofluorescence image of a mouse brain tissue section. \textbf{B} Spatial domains identified by DeepST, SEDR, STAGATE, GraphST, and SemanticST, using Louvain clustering algorithm (resolution=1) on their respective embedding spaces.
    \textbf{C} Allen Reference Atlas of mouse hippocampus from SlideseqV2 dataset, serving as ground truth for comparison.  \textbf{D} Spatial domains derived from STAGATE, GraphST, and SemanticST embedding spaces using Louvain clustering algorithm (resolution=0.3). \textbf{E} Spatial domains identified by SemanticST alongside their corresponding marker genes.
    \textbf{F} Tissue annotations of the E9.5 mouse embryo Stereo-seq dataset from the original study. \textbf{G} Spatial clustering results of the mouse embryo Stereo-seq dataset obtained by GraphST and SemanticST. \textbf{H} Separate visualization of key embryonic organs identified by SemanticST. \textbf{I} Visualization of marker genes characterizing each organ.}
    \end{minipage}
    \noindent\rule{\textwidth}{0.4pt} 
    \vspace{1em}
    \addtocounter{figure}{-1} 
}

To further evaluate whether each semantic graph reveals biological insights, we generated heatmap plots of the learned attention scores for each Semantic graph, as illustrated in Supplementary Figure S4C. Our approach involved first categorizing each spot into one of four groups, based on which semantic graph yielded the highest attention score for that spot. Subsequently, we conducted differential expression analysis among these four classes. The heatmap plots in Supplementary Figure S4D display the top three most highly expressed genes for each semantic graph, providing a visual representation of the distinct gene expression patterns associated with each graph.

Semantic Graph 1 highlights regions predominantly expressing \textit{Nrgn}, \textit{C1ql3}, and \textit{Neurod6}. \textit{Nrgn} is primarily expressed in the forebrain, particularly in the hippocampus and cerebral cortex, where it plays a role in synaptic plasticity and memory processes \cite{zhong2009neurogranin}. Notably, \textit{Nrgn} is also a well-documented schizophrenia risk gene \cite{zhang2019association}, linking this region to potential neuropsychiatric implications. \textit{C1ql3} expression is observed in specific neuronal populations within the forebrain \cite{martinelli2016expression}. \textit{Neurod6} is characteristically expressed in pyramidal neurons of the neocortex and hippocampus \cite{kay2011neurod6}. The expression patterns of these genes collectively suggest that the first Semantic Graph primarily represents forebrain structures, including the hippocampus and cerebral cortex, areas crucial for cognitive functions and potentially implicated in neuropsychiatric disorders like schizophrenia.

We found highly expressed genes like \textit{Apod} (primarily expressed in oligodendrocytes and astrocytes), \textit{Mobp} (specifically expressed in oligodendrocytes), and \textit{Ptgds} (expressed in oligodendrocytes) \cite{zhang2014rna} in the second Semantic Graph, highlighting oligodendrocyte-rich regions, including the thalamus and white matter tracts. The third Semantic Graph is characterized by high expression of genes such as \textit{Vim}, \textit{Ifitm3}, and \textit{Pltp}. The co-expression of these genes, typically associated with astrocytes and other glial cells \cite{zhang2016purification}, suggests that this graph likely represents regions rich in glial cells, particularly astrocytes, and potentially areas involved in neuroinflammatory responses or blood-brain barrier function. \\
We then utilized Slide-seq V2 datasets acquired from the mouse hippocampus and compared the results to those obtained using recent approaches, STAGATE and GraphST. Both methods employ graph neural networks—STAGATE with an attention mechanism and GraphST with contrastive learning. The annotated Allen Brain Atlas served as our ground truth, as shown in Fig.~\ref{figure_mouse_brain}C. Similar to our approach with the Visium dataset, we applied the Louvain clustering algorithm with the same resolution to ensure fairness in comparison (Fig.~\ref{figure_mouse_brain}D). Additionally, we highlighted region-specific marker genes in relation to SemanticST domains (Fig.~\ref{figure_mouse_brain}E).

All methods successfully identified major anatomical regions, such as oligodendrocytes and astrocytes, as shown in Fig.~\ref{figure_mouse_brain}D, corresponding to domains 3 and 0 in SemanticST, respectively. However, STAGATE struggled to fully detect the DG, producing an overly narrow cluster for this region (domain 11). For the CA1 and CA2 regions (domain 5), the clustering result was excessively thick. Additionally, within the CA1 region, STAGATE identified a narrow central cluster but incorrectly delineated a broad, fragmented edge around it, corresponding to domains 6 and 8, which deviated from the expected anatomical boundaries.

Both SemanticST and GraphST performed better in identifying these regions, with SemanticST producing less fragmented clusters than GraphST. SemanticST also better captured the layered structure in the cortex (domains 0, 1, and 8). In terms of visualizing the regions around the third ventricle (V3), medial habenula (MH), and lateral habenula (LH), although both STAGATE and SemanticST merged the MH and LH domains, their clusters for MH aligned more closely with its marker gene, \textit{Nwd2}, compared to GraphST. GraphST's clusters aligned more with the anatomical shape but did not correspond well with the marker gene.

Unlike the other two methods, SemanticST was able to identify V3, and although it did not align perfectly with the anatomical region, it was accurately aligned with the marker gene \textit{Enpp2} (Fig.~\ref{figure_mouse_brain}E).

\subsection{SemanticST reveals precise embryonic organs in mouse embryo ST data using Stereo-seq}
To further evaluate the capabilities of SemanticST, we applied both SemanticST and GraphST to the Stereo-seq dataset from a mouse embryo at E9.5, consisting of 5,913 bins and 25,568 genes, to assess their effectiveness in identifying mouse organs. This dataset originally analysed by Chen et al. \cite{chen2022spatiotemporal} provides a high-resolution spatial map of embryonic organ development (Fig\ref{figure_mouse_brain}F). Chen et al. constructed a spatial graph to encode spatial information, followed by Leiden clustering to identify cluster-specific markers. The original study identified 12 clusters, but since GraphST used 22 clusters to achieve higher resolution clusters, we followed this structure and set the number of clusters to 22. Despite using an increased number of clusters, GraphST's results did not fully align with the tissue annotations (Fig.~\ref{figure_mouse_brain}G). The main discrepancies arose from clusters that generally corresponded to mouse organs but were either too broad, with some regions of the cavity being misclassified as organs like the heart and connective tissue, or clusters that were disproportionately sized, such as the sclerotome, which appeared overly thick, and the dermomyotome, was overly narrow, failing to correspond accurately to their respective embryonic structures. Additionally, GraphST failed to classify the liver as a single cohesive cluster, instead splitting it into two separate clusters.
In contrast, SemanticST generated clusters that more precisely reflected the underlying tissue annotations (Fig\ref{figure_mouse_brain}H and Supplementary Fig.S3C) , with strong alignment to known organ-specific marker genes (Fig\ref{figure_mouse_brain}I and Supplementary Fig.S3D). Notably, SemanticST correctly identified the liver region using markers such as \textit{Afp} and \textit{Alb}, the connective tissue using \textit{Postn}, the sclerotome using \textit{Pax1}, and the dermomyotome using \textit{Myog}. Furthermore, head mesenchyme was identified by \textit{Crym}, while the broader mesenchyme was characterized by \textit{Meox1} and \textit{Pcp4}.
A particularly significant improvement was observed in the classification of the heart. SemanticST successfully delineated the heart as a single functional unit, supported by multiple well-established marker genes, including \textit{Myl7,Myl2, Myh6, Myh7, Tnni3, Nppa}, and \textit{Ttn} (Supplementary Fig.S3D). This clustering outcome aligns well with the tissue annotations from the original study and provides a robust representation of the heart as a single functional unit at this developmental stage. In contrast, GraphST incorrectly identified two separate heart clusters, corresponding to the atrium and ventricle, with \textit{Nppa}, a chamber-enriched marker primarily expressed in the atrium and early ventricle (Supplementary Fig.S3D). While \textit{Nppa} is a known marker reflecting spatiotemporal expression during cardiac chamber development, it stands alone as a compartment-specific marker, whereas the one-cluster solution is supported by a broader set of heart-specific markers. This highlights that the single-cluster solution captured by SemanticST better represents the full spectrum of heart-specific gene expression. Notably, beyond organ-specific accuracy, SemanticST also correctly classified the cavity region surrounding the embryo, distinctly separating it from the developing tissues (red cluster)(Supplementary Fig.S3C). This distinction was poorly handled by GraphST, which produced fragmented and inconsistent cavity clusters. This analysis highlights SemanticST’s ability to provide precise and biologically meaningful delineation of embryonic structures, improving alignment with known anatomical features and marker genes. This underscores its potential as a valuable tool for high-resolution ST analysis in developmental biology.

\begin{figure*}[ht!]
    \centering
    \begin{adjustbox}{max width=\textwidth}
        \begin{tikzpicture}

\node at (10,10) {\includegraphics[scale=1]{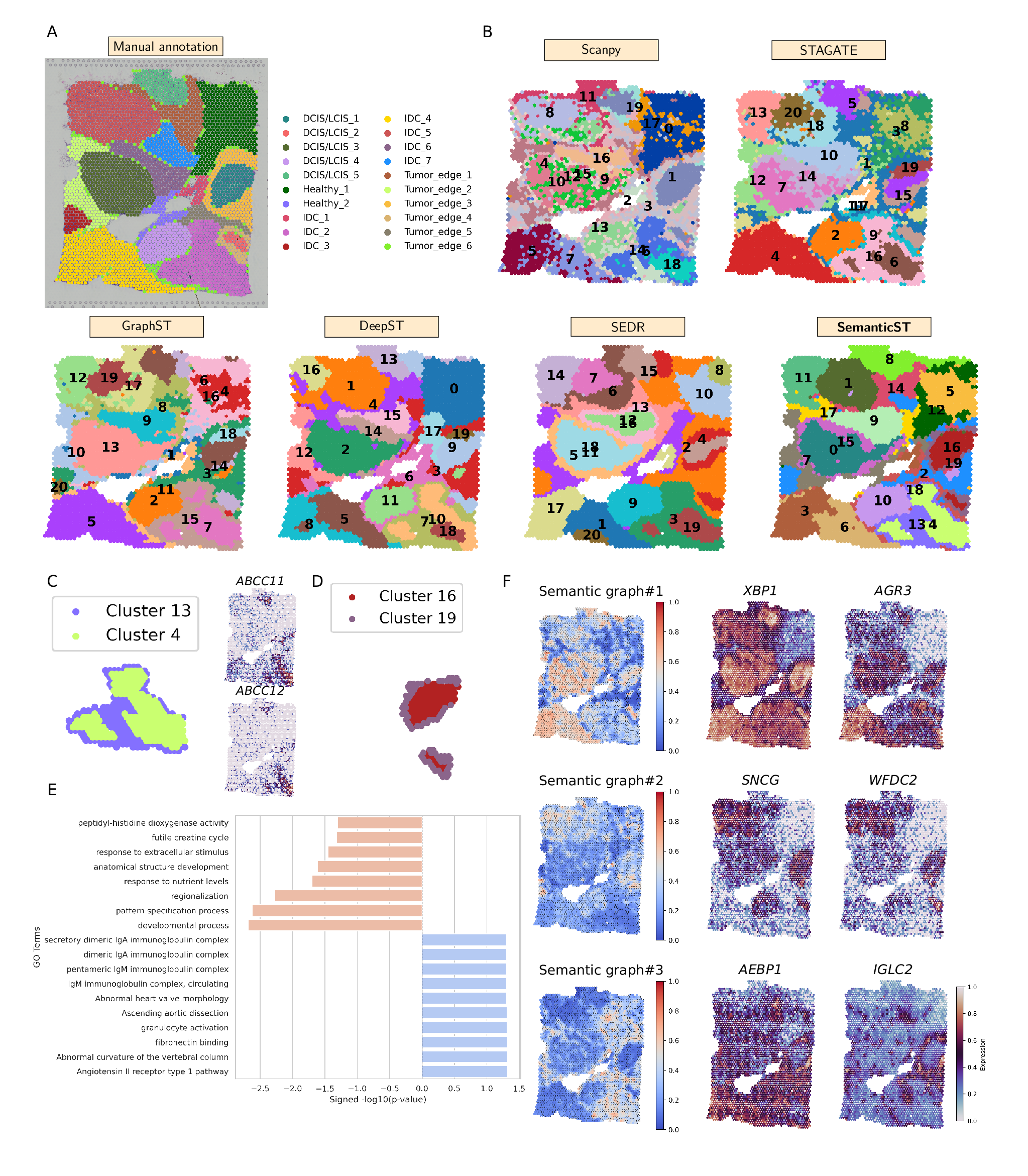}};
        \end{tikzpicture}
    \end{adjustbox}
\caption{Application of SemanticST on Visium ST data of human breast cancer.
\textbf{A.} Manual annotation of breast cancer tissue by SEDR.
\textbf{B.} Clustering results from six different methods.
\textbf{C.} Clusters 13 and 4 detected by SemanticST, along with their associated differentially expressed genes.
\textbf{D.} SemanticST-detected clusters 16 (tumour core) and 19 (tumour edge).
\textbf{E.} GO enrichment analysis for upregulated and downregulated genes in cluster 16 (tumour core).
\textbf{F.} Heatmap of three semantic graphs with learned semantic scores and their associated highly expressed genes.}
 \label{breast_cancer}
\end{figure*}
\subsection{Enhanced tumour heterogeneity dissection on breast cancer by SemanticST }
To further assess the capabilities of SemanticST in capturing fine-grained tissue architecture , we applied it to a spatial transcriptomics dataset from human breast cancer, an exemplar of highly heterogeneous tissue. We analysed a human breast cancer Visium dataset comprising 3,798 spots and 18,246 genes. We set the number of clusters to 20, following the manual annotation provided by SEDR \cite{SEDR} (see Fig.~\ref{breast_cancer}A). The dataset annotation consists of four main categories: invasive ductal carcinoma (IDC), healthy regions, ductal carcinoma in situ/lobular carcinoma in situ (DCIS/LCIS), and tumour surroundings (tumour edge).

We compared SemanticST against other methods, including Scanpy, STAGATE, GraphST, DeepST, and SEDR, focusing on the fidelity of spatial domain detection (Fig.~\ref{breast_cancer}B). While all methods roughly detected clusters that aligned with the manual annotation, there were key differences in their performance. Scanpy, which does not incorporate spatial information, produced noisy clusters with poorly defined domains. STAGATE improved upon this, but its cluster boundaries were fragmented, with noticeable noise. GraphST outperformed the previous methods in spatial preservation but resulted in over-segmentation, particularly within non-malignant regions, where small, disjoint clusters were common .

In contrast, DeepST, SEDR, and SemanticST produced more coherent spatial domains. These methods produced less fragmented clusters that closely matched the manual annotation. However, DeepST generated smooth spatial domains but erroneously split known IDC regions (e.g., IDC\_7) into multiple clusters and failed to resolve the tumour edge, instead producing numerous biologically uninterpretable micro-clusters. SEDR achieved good alignment with the manual annotation, but its tendency to over-extend tumour edge regions led to misclassification of adjacent healthy tissues as malignant or reactive zones, potentially masking critical transition boundaries..

SemanticST outperformed these approaches by producing more contiguous, biologically plausible clusters that better recapitulated the annotated tumour architecture.. Similar to SEDR, it identified additional tumour surroundings as tumour edges, such as Cluster 15 surrounding IDC and DCIS/LCIS regions. Previous studies have shown that these clusters exhibit distinct functional differences, as revealed by pathway analysis \cite{SEDR}. Furthermore, SemanticST revealed two distinct spatial domains within IDC\_2 (Clusters 4 and 13). Differential expression analysis between these clusters revealed \textit{ABCC11} and \textit{ABCC12}, ATP-binding cassette transporters known for their roles in multidrug resistance and tumour aggressivenes \cite{park2006gene} (Fig.~\ref{breast_cancer}C), indicating potential therapeutic implications.

While, SEDR and SemanticST were able to subdivide DCIS/LCIS\_3 (see Fig.~\ref{breast_cancer}A) into a central tumour core and its surrounding edge (Cluster 0 and Cluster 15 in SemanticST, respectively). SemanticST uniquely detected additional tumour edge surrounding IDC\_1 and DCIS/LCIS\_3, by distinguishing Cluster 16 (core) and Cluster 19 (edge) (Fig.~\ref{breast_cancer}D).

To further explore these findings, we performed differential expression analysis between Cluster 16 and Cluster 19, followed by Gene Ontology (GO) enrichment analysis (Fig.~\ref{breast_cancer}E). Interestingly, genes upregulated in the tumour core (Cluster 16) were enriched for immune-related processes, including IgA and IgM immunoglobulin complexes, granulocyte activation, and fibronectin binding, suggesting active immune modulation and tumor-supportive immune cells. In contrast, genes upregulated in the tumour edge (Cluster 19) were associated with developmental processes, pattern specification, and response to nutrient levels, highlighting the dynamic and adaptive nature of the tumour edge as it facilitates invasion and interaction with the surrounding stroma.

These findings are consistent with previous studies demonstrating that tumour cores exhibit immune-related and differentiation-associated activity, which may suggest immune evasion or immune exhaustion in the core environment. In contrast, tumour edges are enriched for metabolic and translational pathways, supporting tumour invasion and expansion \cite{arora2023spatial}. This highlights the capability of SemanticST to distinguish different spatial domains within tumours and capture the functional heterogeneity between the core and edge, revealing their distinct roles in cancer progression.\\

Finally, we visualised a heatmap of the learned semantic scores corresponding to each semantic graph (Fig.~\ref{breast_cancer}F). We presented three out of four semantic graphs, as the last one served as a reference for the others, as we previously discussed in the DLPFC section. After identifying DEGs for each semantic graph, we observed that Semantic Graph 1 exhibited high expression of \textit{XBP1}, a key driver of tumour progression and metastasis \cite{chen2020emerging}, and \textit{AGR3}, a potential biomarker for early detection of breast cancer \cite{de2023prognostic}. This suggests that Semantic Graph 1 preferentially captures tumour-associated regions.
Semantic Graph 2 was characterised by high expression of \textit{SNCG} and \textit{WFDC2}. \textit{SNCG} (breast cancer-specific gene-1) is an established aggressive marker in breast cancer, while \textit{WFDC2} is indeed an FDA-approved ovarian cancer biomarker \cite{james2020biomarker}, its role in breast cancer is less defined and not widely accepted as a validated biomarker \cite{perou2000molecular}. These findings highlight that Semantic Graph 2 captures a molecularly distinct tumour region.
Lastly, semantic Graph 3 exhibited high expression of immune-related genes, particularly \textit{IGLC2}, an immunoglobulin gene used as a prognostic marker in breast invasive carcinoma \cite{uhlen2015tissue}. Additionally, \textit{AEBP1} was highly expressed in this region, which has been known to promote tumour proliferation, invasion, and resistance to apoptosis in breast cancer cells \cite{li2023aebp1}. This suggests that semantic Graph 3 represents immune-associated and tumour-proliferative regions.
Together, these results demonstrate that SemanticST not only improves tumour heterogeneity dissection but also provides deeper biological insights into the molecular composition of tumour subregions.

 \subsection{SemanticST Reveals the Tumour Microenvironment of Human Breast Cancer in High-Resolution Xenium Data}
In our final analysis, we evaluated the performance of SemanticST on a high-resolution SRT dataset generated by 10x Xenium from human breast cancer \cite{janesick2023high}. 
 Due to the substantial computational intensity and scale of this dataset, we were only able to apply IRIS \cite{IRIS} and Banksy \cite{singhal2024banksy} to this dataset, both of which are non-deep neural network-based methods (Fig.~\ref{figure_XENIUM}A). Other deep learning-based approaches were unable to process this dataset due to its large scale (e.g., most methods cannot handle datasets with more than $90,000$ samples). The original annotations by Janesick et al.  \cite{janesick2023high} (Fig.~\ref{figure_XENIUM}A, and Supplementary Fig.S7) were derived by integrating complementary single-cell data, followed by subclustering of the spatial transcriptomic map. 

While this provides a biologically guided reference, it introduces challenges for direct one-to-one comparison, particularly given that the Xenium gene panel is curated and limited, with emphasis on tumour and immune-associated genes. Thus, our comparison focused on biological interpretability: we used the Janesick annotation as a reference and validated SemanticST-derived domains through enrichment analysis using known lineage and functional marker genes.
 
\subsubsection{Immune, Endothelial, Stromal, and Adipocyte Domains  }
SemanticST accurately identified key immune and stromal cell populations consistent with prior annotations. Domain 0 represents CD4 T-cells, marked by highly expression of \textit{IL7R}, and \textit{CD4} (Supplementary Fig.S6C), while domain 2 corresponded to endothelial cells. Domain 15 captured IRF7+ dendritic cells, involved in antigen presentation and antiviral responses. Furthermore, SemanticST identified segregated domains 1, 6, 12, and 13 as stromal cells, which were highly expressed by \textit{POSTN}, \textit{SFRP1}, \textit{FBLN1}, and \textit{LUM}, and domain 9 as GJB2+ stromal cells (Supplementary Fig.S6A). Additionally, domain 17 enriched for \textit{APOC1}, \textit{ITGAX}, and \textit{CD68} (Supplementary Fig.S6B), corresponded to macrophages, likely reflecting M1-polarised, pro-inflammatory activity. Despite challenges related to adipocyte segmentation due to the lipid-rich content and edge-skirted transcript expression \cite{janesick2023high}, SemanticST was able to identify domain 14, which is highly expressed for \textit{ADIPOQ}, \textit{LPL}, and \textit{MEDAG}, potentially representing adipocyte cells (Supplementary Fig.S6D).

\subsubsection{Myoepithelial, DCIS, and Invasive Tumor Domains }
\begin{figure*}[!p]
    \centering
    \begin{adjustbox}{max width=\textwidth}
        \begin{tikzpicture}

\node at (10,10) {\includegraphics[scale=1]{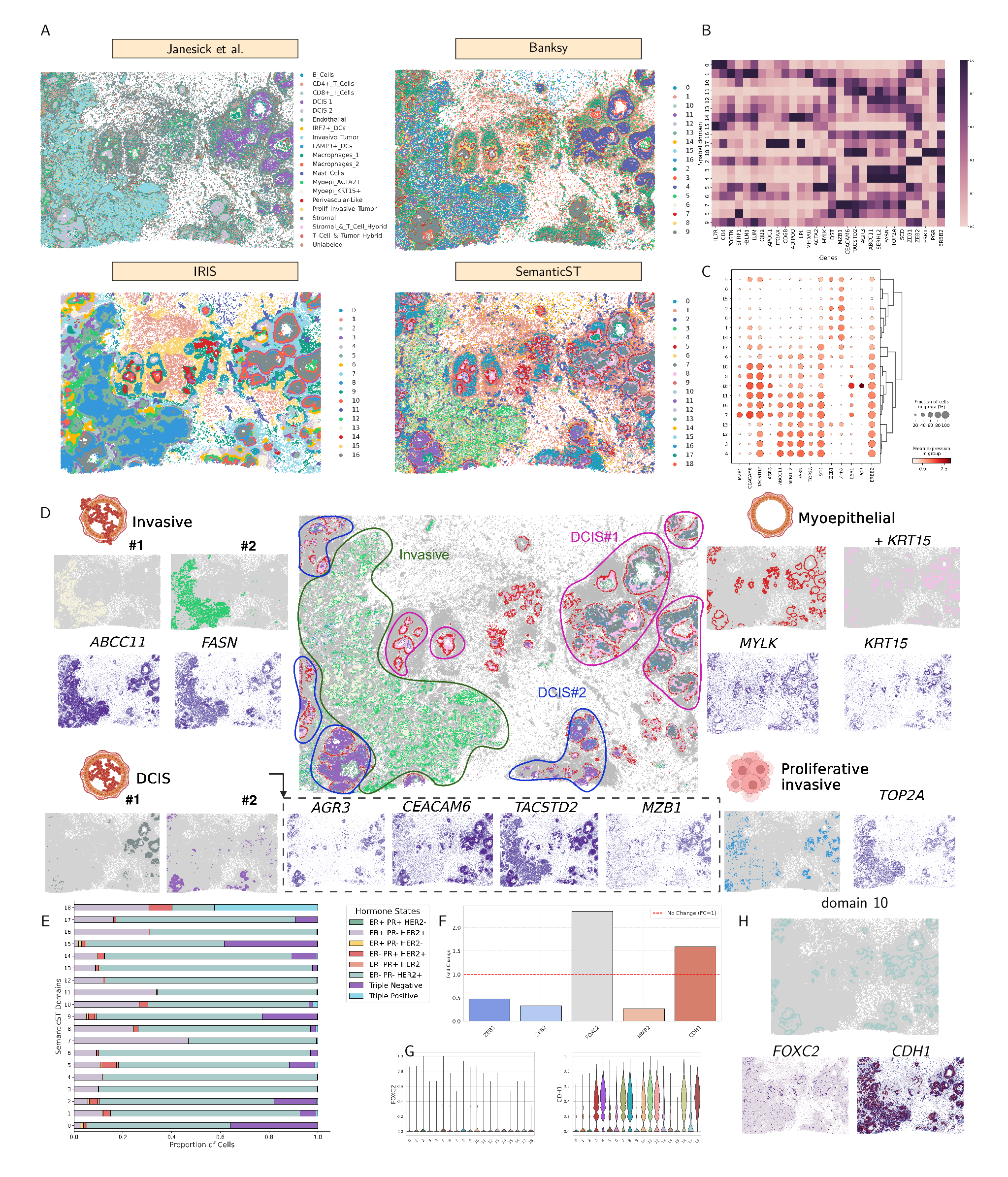}};

        \end{tikzpicture}
    \end{adjustbox}
    \caption{Xenium ST analysis of Breast Cancer. \textbf{A.} Spatial domain identification using data from Janesick et al. (annotation), Banksy, IRIS, and SemanticST. \textbf{B.} Heatmap displaying well-known genes across all spatial domains identified by SemanticST. \textbf{C.} Dot plot illustrating the expression of cancer-related genes across all spatial domains detected by SemanticST. \textbf{D.} Tissue classification based on SemanticST-detected domains of myoepithelial, DCIS, invasive, and proliferative invasive, with associated highly expressed genes. \textbf{E.} Bar plot showing the distribution of hormone receptor statuses within each spatial domain identified by SemanticST. \textbf{F.} Statistical significance of EMT-related genes between domain 10 and other domains, assessed using the Mann-Whitney U test. \textbf{H.} Spatial plots of the significant genes \textit{FOXC2} and \textit{CDH1} in domain 10.}

    \label{figure_XENIUM}
\end{figure*}

Ductal carcinoma in situ (DCIS) is a non-invasive proliferation of epithelial cells confined within the breast ducts, restricted by an intact myoepithelial layer and basement membrane \cite{wilson2022ductal}. In contrast, invasive ductal carcinoma (IDC) occurs when these epithelial cells escape the ductal confinement, break through the myoepithelial layer and basement membrane, and invade the surrounding tissue \cite{wilson2022ductal}. Fig.~\ref{figure_XENIUM}D illustrates these spatial domains, highlighting the myoepithelial, DCIS regions, and both invasive and proliferative invasive areas detected by SemanticST, along with their associated highly expressed genes.

While, all evaluated methods distinguished DCIS from invasive tumour regions, their precision in resolving degree of invasiveness varied. For example, the original study identified two distinct DCIS regions based on invasiveness, attributed to a layer of myoepithelial+KRT15 surrounding DCIS 1 and absent from DCIS 2 (Supplementary Fig.S7). Banksy and IRIS identified this layer as domains 12 and 14, respectively, while SemanticST more accurately aligned with \textit{KRT15} expression and more accurately reflected the degree of invasiveness (Supplementary Fig.S8). Both Banksy and SemanticST successfully detected two distinct clusters for the two DCIS regions, but IRIS grouped them into a single cluster (domain 16).

The Myoepithelial compartment (domains 5, 10, and 7 in SemanticST, IRIS, and Banksy, respectively) is expected to surround DCISs based on the expression of marker genes like \textit{ACTA2}, \textit{MYLK}, and \textit{DST}. However, IRIS was unable to accurately detect this domain (Supplementary Fig.S8).

SemanticST successfully identified cluster 7 as DCIS 1, showing increased expression of \textit{MZB1}, a key distinguishing marker for DCIS 1 in the original study, \textit{CEACAM6}, a cell adhesion molecule \cite{han2008ceacam5}, and \textit{TACSTD2}, a tumour-associated epithelial marker (Fig.~\ref{figure_XENIUM}D). Additionally, SemanticST identified cluster 11 as DCIS 2, showed high expression of \textit{CEACAM6} and \textit{AGR3}, both linked to early-stage epithelial transformation. 
SemanticST identified domains 3 and 4 as invasive tumour regions, characterized by the high expression of \textit{ABCC11}, \textit{SERHL2}, and \textit{FASN}, and the absence of a myoepithelial layer. Additionally, domain 16 was marked by high expression of  \textit{TOP2A} and \textit{SCD} genes associated with tumour proliferation and metabolic activity, which can be clustered separately (Fig.~\ref{figure_XENIUM}D). Despite the significant spatial scattering of this region, SemanticST successfully segregated this domain.

\subsubsection{Characterization of a Small Triple Receptor-Positive Region}
Hormone receptor status, defined by estrogen (ER/\textit{ESR1}), progesterone (PR/\textit{PGR}), and the amplification of human epidermal growth factor receptor 2 (HER2/\textit{ERBB2}), can influence breast cancer classification and treatment strategies. Although the dataset used in this study was annotated as HER2/+/ER+/PR− by a pathologist, SemanticST uniquely identified cluster 18—a small cluster with co-expression of \textit{PGR}, \textit{ERBB2}, and \textit{ESR1} (Supplementary Fig.S6E), revealing a rare triple receptor-positive region with potential therapeutic relevance. Notably, SemanticST is the only method capable of detecting this biologically distinct region despite its small size. Banksy consider this region as DCIS 1 (cluster 4), and IRIS detected inside cluster 10 which represent myoepithelial cells. We also carefully analysed the status of the three receptors across all domains to better characterise tumour heterogeneity. Subsequently, Fig.~\ref{figure_XENIUM}E shows the distribution of spots with specific hormone receptor statuses (\textit{ER, PR, HER2}) across spatial domains detected by SemanticST.

\subsubsection{Tumor-Associated Myoepithelial Cells in the DCIS to IDC Transition}
Since the transition from DCIS to IDC is not fully understood and remains an area of ongoing research  \cite{janesick2023high}, identifying the microenvironment of each region is essential for furthering our understanding.
While normal myoepithelial cells are traditionally considered to play a tumor-suppressing role, recent studies have highlighted the presence of tumor-associated myoepithelial cells as a rare cell type with tumor-promoting effects \cite{lo2017tumor}.
A recent study \cite{lo2017tumor} explored this phenomenon and suggested that $TGF\beta$ signalling pathway contributes to DCIS progression by promoting the development of DCIS-associated myoepithelial cells.

In contrast to the original study by Janesick et al., which delineated cell boundaries using subclustering of epithelial and myoepithelial cells along with scFFPE-seq data, SemanticST identified domain 10, as a boundary-like region that harbours both tumor-associated markers (\textit{CEACAM6, TACSTD2}) and myoepithelial markers (\textit{MYLK, DST}) (Supplementary Fig.S6F). 
Since the $TGF\beta$ signaling pathway is known to induce epithelial-mesenchymal transition (EMT), we investigated the expression of several EMT-related genes in this dataset, including \textit{ZEB1}, \textit{ZEB2}, \textit{FOXC2} , \textit{MMP2}, and \textit{CDH1}, across all domains. We specifically focused on whether their expression differs in domain 10. To do that, we applied the Mann–Whitney U test \cite{mcknight2010mann}, a non-parametric statistical test suitable for gene expression data (Fig.~\ref{breast_cancer}F).
\textit{ZEB1} and \textit{ZEB2} were significantly downregulated in domain 10 (fold change = 0.47, $p$ = 1.03e-136; fold change = 0.33, $p$ = 0.00), indicating a lack of canonical EMT activation. In contrast, \textit{FOXC2} was significantly upregulated (fold change = 2.34, $p$ = 1.14e-92), suggesting the activation of a non-canonical EMT programme. Despite the presence of EMT regulators, \textit{MMP2} was downregulated (fold change = 0.26, $p$ = 0.00), indicating incomplete mesenchymal transition. \textit{CDH1}, an epithelial marker, was higher in domain 10 (fold change = 1.59, $p$ = 0.00), suggesting retention of epithelial traits. The expression of \textit{CDH1} and \textit{FOXC2} across all domains, along with their spatial expression, is illustrated in Fig.~\ref{breast_cancer}G and ~H, respectively. These findings indicate that domain 10 exhibits partial EMT, with non-canonical EMT activation and preserved epithelial characteristics. 
This domain—undetected by the original study Janesick et al. as well as Banksy and IRIS—corresponds to tumour-associated myoepithelial cells and aligns with the findings of Lo et al \cite{lo2017tumor}. Importantly, we observed the higher expression of \textit{FOXC2} in this domain as well as myoepithelial domain (domain 5). Beyond its role in EMT regulation, \textit{FOXC2} expression has been linked to stem-like cancer cell states and may represent a potential a therapeutic target  \cite{hollier2013foxc2}.

\subsection{SemanticST Enables Vertical and Horizontal Integration of Multiple Samples by Mitigating Batch Effects}
\begin{figure*}[ht]
    \centering
        \centering
            \begin{adjustbox}{max width=\textwidth}
        \begin{tikzpicture}

        \node at (10,10) {\includegraphics[scale=1]{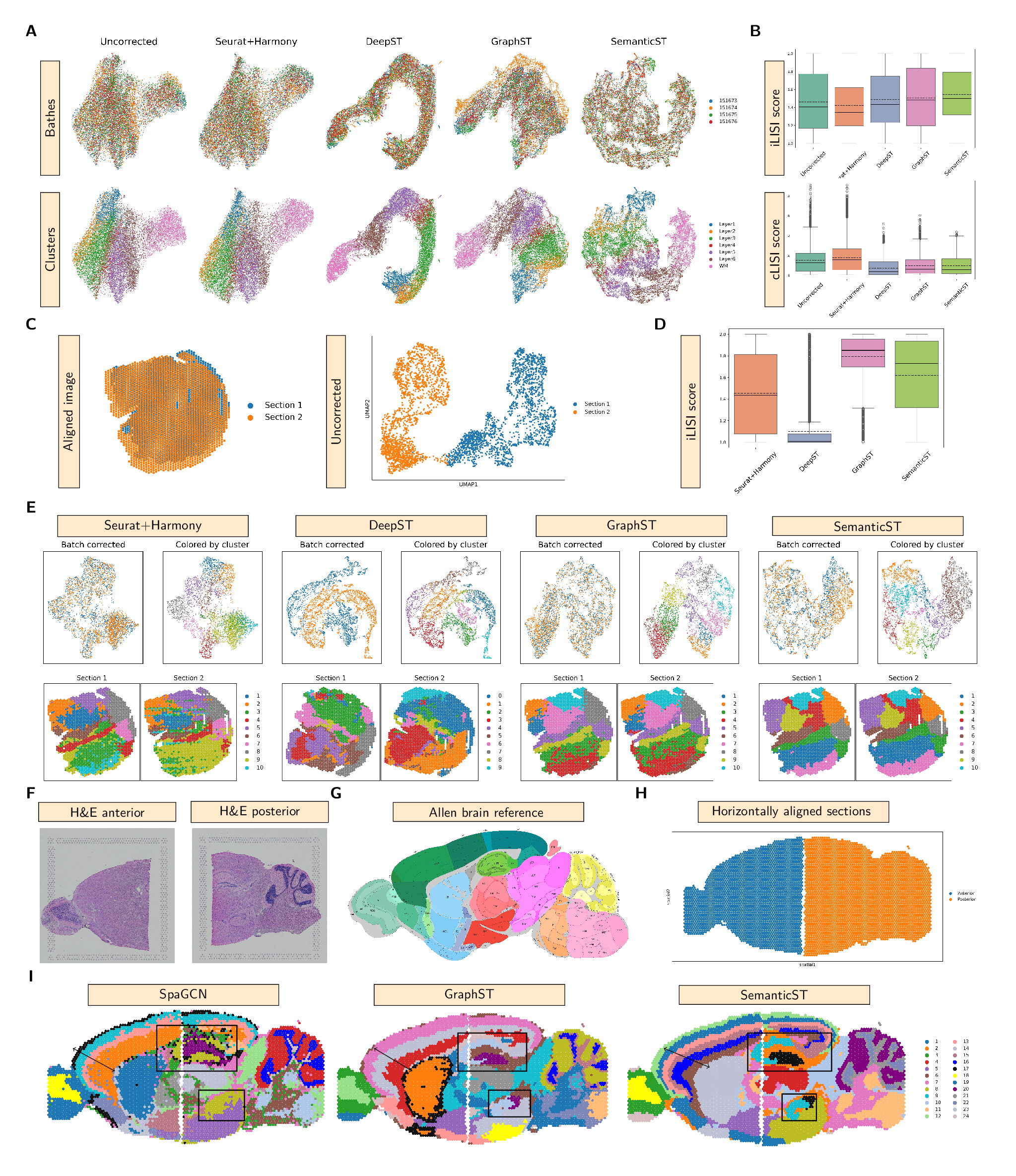}};
        
        \end{tikzpicture}
    \end{adjustbox}
   \phantomcaption
    \label{figure_Integration}
\end{figure*}

Lastly, we evaluated SemanticST’s ability to jointly embed multiple tissue slices while correcting for batch effects, focusing on both vertical and horizontal integration. Vertical integration poses greater challenges due to stronger batch-induced variability. An effective integration method should minimize technical artifacts while preserving biological heterogeneity. We first applied SemanticST on four consecutive slices from DLPFC datasets, including sections \#151673, \#151674, \#151675, and \#151676, and compared it to Seurat+Harmony, DeepST, and GraphST. UMAP plots Fig.~\ref{figure_Integration}{A} show batch assignment (top) and biological clusters (bottom).

The uncorrected data exhibited pronounced batch effects. Seurat+Harmony reduced these effects but retained minor batch-specific patterns. DeepST and GraphST improved integration but showed residual batch separation. SemanticST outperformed all, displaying minimal batch-specific clustering and the best overall integration. 
We quantitatively assessed integration using iLISI (integration local inverse Simpson’s Index) and cLISI (corrected local inverse Simpson’s Index). As showed in Fig.~\ref{figure_Integration}B, SemanticST achieved the highest median iLISI and a low-variability cLISI score, indicating consistent batch correction and preservation of meaningful structure. For cLISI (lower is better), all methods except Seurat+Harmony showed similar median scores. DeepST had the lowest median but notable variability, as did GraphST. SemanticST had a slightly higher median cLISI but the narrowest IQR, indicating more consistent and reliable performance. 

Next, we evaluated performance on a mouse breast cancer dataset (two 10x Visium sections). The UMAP plots (Fig.~\ref{figure_Integration}C) revealed batch effects between sections. Since no ground truth was available, we used the iLISI score to assess the quantitative results (Fig.~\ref{figure_Integration}D). Additionally, Fig.~\ref{figure_Integration}E shows the UMAP plot of the data in the first row and the corresponding clustering results in the second row. Seurat+Harmony slightly mitigated batch effects, achieving a higher iLISI score than DeepST. Although the clusters between the two sections were aligned, they appeared fragmented without less well-defined separations. DeepST had the weakest integration result, with poor batch-effect removal, the lowest iLISI score, and notable outliers, as well as dissimilar clusters across sections. While GraphST outperformed other methods in terms of the iLISI score, it also showed a considerable number of outliers, leading to inconsistent results. In the GraphST clustering results, the clusters were aligned between the two sections, but significant fragmentation was still evident, especially between clusters 3 (green) and 4 (red). SemanticST demonstrated the best post-integration performance, with consistent quantitative results, a slightly lower iLISI score than GraphST but without outliers, and well-aligned clusters between the two sections without fragmentation. \\

\afterpage{
    \clearpage
    
    \noindent\rule{\textwidth}{0.4pt} 
    \captionsetup[figure]{labelformat=empty} 
    \begin{minipage}{\textwidth}
    \captionof{figure}[Example Figure]{\textbf{Fig.6 |}
    
 \textbf{SemanticST offers precise vertical integration of ST data from the human cortex and mouse breast cancer, as well as horizontal integration of data from the anterior and posterior regions of the mouse brain.} \textbf{A} UMAPs displaying the integration results of four methods (Seurat+Harmony, DeepST, GraphST, and SemanticST) compared to uncorrected data across four sections of the human dorsolateral prefrontal cortex (DLPFC) dataset, specifically sections \#151673, \#151674, \#151675, and \#151676. \textbf{B} iLISI and cLISI scores for the integration results of these four DLPFC sections using the latent representations generated by the four methods (Seurat+Harmony, DeepST, GraphST, SemanticST) and the uncorrected data. \textbf{C} Aligned image of mouse breast cancer using the PASTE algorithm, alongside a UMAP plot of the PASTE output prior to batch correction. \textbf{D} Boxplots of iLISI scores for mouse breast cancer across the four methods. \textbf{E} UMAPs after batch effect correction and clustering obtained by Seurat+Harmony, DeepST, GraphST, and SemanticST, respectively. The second row shows clustering results for two sections separately, based on the latent representation of each respective method. \textbf{F} Histology images of the anterior and posterior regions of mouse brain section 1. \textbf{G} Mouse brain section annotated using the Allen Mouse Brain Atlas as ground truth. \textbf{H} Horizontal integration of two sections. \textbf{I} Clustering results on the latent representation of SpaGCN, GraphST, and SemanticST for the horizontally integrated sections of mouse brain section 1.}
    \end{minipage}
    \noindent\rule{\textwidth}{0.4pt} 
    \vspace{1em}
     \addtocounter{figure}{-1} 
}

For horizontal integration or slice alignment, we evaluated SemanticST on two horizontally wrapped slices of anterior and posterior sections (Fig.~\ref{figure_Integration}F) from mouse brain 10x Visium data. Our aim was to assess its ability to jointly analyze multiple slides and compare the results to the mouse brain atlas annotation \cite{atlas2006allen} (Fig.~\ref{figure_Integration}G). Since these slides are adjacent, we aligned the spatial coordinates by adjusting and scaling them to ensure horizontal alignment, followed by concatenating the datasets for joint analysis (Fig.~\ref{figure_Integration}H). As SpaGCN was the first method to perform joint analysis on this dataset, we compared SemanticST against SpaGCN and GraphST, using the same number of clusters (24) as detected by SpaGCN through Louvain clustering. For a more detailed comparison, we focused on the hippocampal (highlighted by the black rectangle) and cortex regions (black arrows). As shown in Fig.~\ref{figure_Integration}I, all methods successfully identified clusters that aligned with the reference atlas, detecting most anatomical regions, particularly the vermal region. However, while SpaGCN detected the dorsal horn of the hippocampus, it missed the ventral part (bottom rectangle), and most clusters appeared fragmented, resembling noisy clusters. GraphST and SemanticST performed better, both detecting hippocampal regions across tissue sections. Nevertheless, GraphST showed some degree of fragmentation in its clusters and, like SpaGCN, identified only three of the five cortex layers. In contrast, SemanticST outperformed both methods by accurately identifying all cortex layers, providing smoother, more aligned anatomical regions, and detecting clusters across both slides without fragmentation.

\section*{Discussion}
Spatial transcriptomics offers a transformative lens into tissue organization, enabling the study of cellular heterogeneity within its native microenvironment. In this study, we propose SemanticST, a graph neural network-based unsupervised deep learning approach for ST data analysis. Importantly, ST is incredibly complex, and analysing it like other types of data can introduce biases. These biases often allow dominant patterns to overshadow crucial but subtle ones. Rare cell types or small differences in gene expression, especially in cancer, can have profound consequences. These small variations might seem insignificant at first but can lead to critical outcomes, making it essential to carefully consider them in analyses \cite{trapnell2015defining}. Therefore, inspired by the complexity of biological systems, we introduce the concept of a semantic graph to represent the intricate relationships between transcriptomes across tissue.
SemanticST was developed to address the complexity of biological systems through a fundamentally different design. Rather than using a single, spatially constrained graph, our method employs a dynamic MLP-based attention mechanism to construct multiple semantic graphs, each capturing distinct biological relationships beyond spatial proximity. Unlike other attention-based methods \cite{STAGATE}, which apply a shared attention mechanism across a single spatial graph, our method learns separate attention scores for each semantic space. This attention-based graph construction enables the model to extract rich, disentangled representations from heterogeneous tissues. Moreover, SemanticST leverages the min-cut loss function as an alternative to contrastive learning methods \cite{graphST}, which depend on the careful selection of positive and negative samples. This allows the model to learn flexible representations without reliance on noisy sampling procedures or post hoc graph corrections.
We generated heatmaps of attention scores corresponding to each semantic graph for Visium data, providing biological validation for each extracted graph. Notably, we do not claim that semantic graphs can capture region-specific cancer types or mental disorders. However, while semantic graphs may highlight regions associated with specific diseases (e.g., mental health disorders or breast cancer in different datasets), this is not a predefined function of our model. Rather, our method captures biologically meaningful structures in the data, and clustering further helps reveal these specific regions. We state that our proposed method functions as a kind of explainable AI, providing a logical flow for decision-making.
We applied SemanticST to various ST datasets with different resolutions, ranging from 10x Visium to Xenium. The results demonstrated the superiority of SemanticST over recent state-of-the-art methods. Additionally, SemanticST showed exceptional performance in ST data integration and batch effect removal, as evidenced by our results. 
In our study, we analysed two distinct breast cancer tissue samples profiled using Visium and Xenium ST, motivated by breast cancer’s complex pathogenesis and substantial mortality rate \cite{siegel2024cancer}. This integrative analysis enabled us to capture both the broad transcriptional landscape and fine-grained spatial features across different technological platforms, contributing to a more comprehensive understanding of tumour heterogeneity.
In particular, in the Visium sample, where a larger number of genes are available as feature dimensions, we were able to precisely analyse both the tumour and tumour edges. Leveraging SemanticST, we were able to accurately identify even the smallest spatial domains, including subtle gene expression patterns along the tumour edges. This fine-grained ability to detect transitional tumour regions may provide valuable insight into tumour-stroma interactions and early invasion mechanisms.
To test its resolution on higher density ST, we analysed the Xenium sample from Janesick et al. (2024) \cite{janesick2023high}. Their study successfully identified DCIS and other regions by integrating three types of data from a sample (including Xenium, Visium and scFFPE-seq), which emphasized the necessity of an integrative approach to gain meaningful insights. Importantly, we demonstrate that our approach, relying solely on the Xenium data, is equally effective. While their study found that clustering the Xenium data alone, without the use of external platforms, failed to distinguish the two distinct DCIS cell types, we were able to uncover these regions with just the Xenium data using SemanticST. This highlights the power of our method, which bypasses the need for additional technologies or multiple samples, ultimately saving both cost and time. SemanticST could accurately identify DCIS regions with different degrees of invasiveness based on the myoepithelial layer and correctly distinguish small spatial domains of the three positive hormones \textit{ER, PR, HER2} as separate clusters. It also detected a tumor-associated myoepithelial domain highly expressed by \textit{FOXC2}, a gene well-known for its central role in the regulation of EMT and metastasis. These FOXC2+ niches may reflect a partial EMT state with stem-like and invasive properties, supporting recent models of phenotypic plasticity in breast cancer.
A particularly notable strength of SemanticST is its ability to process ultra-large datasets. To our knowledge, it is the first deep learning model capable of analysing Xenium data at cellular resolution while incorporating spatial information. Competing methods like IRIS and Banksy either ignore learning-based representations or rely on heuristic graph construction, limiting their capacity to model non-linear transcriptional patterns. SemanticST overcomes these barriers through efficient mini-batch training and encoder-based embeddings, ensuring both scalability and biological expressiveness, even when gene panels are sparse, as in targeted ST platforms.
In most datasets, SemanticST considers the decoder output as the final latent representation, a smoothed version of gene expression that incorporates spatial context and semantic structure. However, for Xenium, which contains a very large number of spatial spots but relatively few genes, this reconstruction step becomes overly constrained. The decoder is inherently under-parameterised in such low-dimensional settings and may tend to overfit to noise. Hence, we use the encoder output as the final embedding for downstream analysis in Xenium.
Another key output of SemanticST is the semantic score, learned through the graph learning process. Since each spot or cell receives a semantic score per graph, these embeddings could serve as a foundation for exploring biologically informed cell–cell or spot–spot interactions, a direction we plan to explore further. While SemanticST focuses on transcriptomic data, future extensions should integrate spatial proteomics, chromatin accessibility, and imaging data to capture a fuller picture of the tissue state. We also acknowledge that cell-type deconvolution was not pursued, as many new ST platforms operate at single-cell resolution, reducing its immediate relevance.
Looking ahead, we envision SemanticST as a foundational tool for building biologically grounded, clinically useful representations of tissue architecture. Its scalability, interpretability, and cross-platform compatibility position it as a core component in future spatial multi-omics pipelines. Future directions include adaptation to ISH-based platforms, incorporation of segmentation modules, and use as a pretraining framework for spatial foundation models. As spatial omics technologies continue to evolve, models like SemanticST will play an essential role in translating spatial complexity into actionable biological and clinical insights.
\section*{Method}
\subsection{Data preprocessing}
SemanticST uses a gene expression matrix and spatial coordinates as input. Initially, genes expressed in fewer than 50 spots or cells are filtered out, and cells expressed in fewer than 20 genes are also excluded. The raw counts are then log-transformed and normalized by library size using the SCANPY package \cite{wolf2018scanpy}. The gene and cell filtering steps are dependent on the specific ST technology being used. For instance, with Visium-based ST data, where each spot contains multiple cells, the filtering steps are not applied. For Xenium data, gene filtering is ignored, and for all other data types, the top 3000 most variable genes are selected as the input for SemanticST.
\subsection{Graph neighbourhood construction}
To capture the spatial dependencies, which are a key advantage of ST data, we model these dependencies using an undirected graph. This graph, denoted as $G(V,E)$, consists of a set of vertices $V$, where each $v \in V$ represents a spot or cell, and edges $E$ that define the connections between them. The graph is constructed by computing the pairwise Euclidean distance between spots or cells and connecting each node to its $k$ nearest neighbours. From this graph $G$, we derive the adjacency matrix $A \in \mathbb{R}^{ N_{spot/cell} \times N_{spot/cell}}$, where $ N_{spot/cell}$ represents the total number of spots or cells. Each entry $a_{ij}$ in $A$ is set to 1 if spots or cells $i$ and $j$ are neighbours, and 0 otherwise. To incorporate self-loops, we define the modified adjacency matrix $\hat{A}=A+I$, where $I$ is the identity matrix. The degree matrix ${D}$ is a diagonal matrix with elements defined as ${D}_{ii}=\sum_{j=1}^{N_{spot/cell}} \hat{A}_{ij} $, with $\hat{A}$ being the adjacency matrix that includes self-loops. Subsequently, the normalized adjacency matrix is given by $\tilde{A}=D^{\frac{-1}{2}}\hat{A}D^{\frac{-1}{2}}$. In the SemanticST implementation, we set $k=5$, as this value provided the best performance in our analysis. 
\subsection{SemanticST}
SemanticST introduces an unsupervised graph neural network that constructs a semantic graph to learn latent representations of ST data. These embeddings enable tasks such as spatial clustering, data integration, and other downstream analyses. Figure~\ref{fig:semantic} illustrates the complete SemanticST pipeline, which consists of four key steps: (i) Semantic graph learning, (ii) GNN-based encoder-decoder for ST embedding learning, (iii) Fusion of latent representations across semantic graphs, and (iv) Community-based detection loss function. Each step is explained in the following sections.
\subsubsection{Semantic graph learning}
We hypothesize that there exist higher order relationships among samples, suggesting the presence of multiple semantic relation spaces. Each of these spaces captures distinct aspects of the data and can be represented as weighted graphs reflecting the structure of the input graph. Drawing inspiration from the Graph-Mixed-up method \cite{wu2021graphmixup}, we aim to extract disentangled semantic features using a semantic feature extractor. This extractor takes into account not only Euclidean spatial relations but also other relation spaces between spots and cells. Firstly, we transfer the preprocessed gene expression matrix \( X \in \mathbb{R}^{N \times F} \) with $N$ nodes and $F$ features per node to a low-dimensional space \(H \in \mathbb{R}^{N \times D_{f}} \) as follows:
\begin{equation}
    H = XW_h,
\end{equation}
where \(W_{h} \in \mathbb{R}^{F \times D_{f}}\) is a parameter matrix. Then, we use the transformed feature $H$ to generate \(k(1 \leq k \leq K)\) semantic graphs as follows:
\begin{equation}
G_k = \sigma\left(W^{k}_{m} \cdot  (h_i\oplus h_j)  + b^{k}_{m}\right).
\end{equation}
Here, \(\sigma=Sigmoid(.)\) represents the activation function applied to the output of a one-layer multi-layer-perceptron (MLP) $m$, which takes the transformed features of nodes $i$ and $j$ as inputs, and $\oplus$ is the vector concatenation operation. $W_{m}$ and $b_m$ denote a trainable weight and bias vector, respectively. The function calculates the relationship between each pair of nodes, which we consider as the weight of the edge connecting these nodes in the graph. The output of the MLP functions as a semantic graph, which is essentially a weighted graph maintaining the same structure as the input graph to preserve spatial dependencies. The weights in this graph represent the relationships between spots or cells in the high-dimensional space. 
Since each semantic graph captures a unique aspect of the data, a graph descriptor $d_{k}$ is used to map low-dimensional inputs and semantic graphs into a vector representation. This approach is essential because, by applying constraints to the learned semantic graphs, we enable the model to generate different graphs that collectively capture the complexity of the ST data.
Using $G_k$ as the weight of the graph and $H$ as the feature matrix of the nodes, we employ an encoder $s$ consisting of two GCN layers, as expressed by the following formula:
\begin{equation}
Z_s^{(l,k)} = \sigma\left( G_k \cdot \tilde{A} \cdot W_{s}^{(l,k)} \cdot Z^{(l-1)} \right).
\end{equation}
Here, $Z_s^{(l,k)}$ represents the feature representation of semantic graph $k$ at layer $l$, while $Z_s^{(0,k)}$ is initialized as the transformed gene expression matrix $H$. The matrix $W_{s}^{(l,k)}$ is the learnable weight matrix of encoder $s$ for the $l$-th layer of semantic graph $k$, and $\sigma$ refers to the activation function, which is the ReLU function. 

Following this, a global average pooling operation, denoted as $\text{Readout}(\cdot)$, is applied for aggregation. Subsequently, a graph descriptor for each semantic graph is computed using a fully connected layer $f$, as follows:
\begin{equation}
d_k = \text{Readout}(Z_s^{(k)}) \cdot W_{f}^k + b_{f}^k.
\end{equation}
In this formula, $W_{f}^k$ and $b_{f}^k$ represent the learnable weight and bias matrices of the fully connected layer $f$.

To optimize the model parameters, including [$W_h, W_m, b_m, W_s, W_f,b_f$] for each semantic graph, we employed the loss function $L_{dis}$ that trains the weight parameters of the descriptor, MLP, and the feature extractor that operates on low-dimensional data. $L_{dis}$ is defined as:
\begin{equation}
L_{dis} = \sum_{i=1}^{K-1} \sum_{j=i+1}^{K} \frac{\mathbf{d}_i \cdot \mathbf{d}_j^T}{\|\mathbf{d}_i\| \|\mathbf{d}_j\|}.
\end{equation}
In order to promote the homogeneity of graph's weights across the neighbouring nodes, while emphasizing the heterogeneity of non-neighbouring nodes, we propose a node similarity loss component, called $L_{node}$. This loss forces the characteristics of neighbouring nodes in the graph to be more uniform. $L_{node}$ is defined as:
\begin{equation}
L_{\text{node}} = \sum_{k=1}^{K} \sum_{i=1}^{N} \sum_{j \in \mathcal{N}(i)} \|\mathbf{d}^{k}_i - \mathbf{d}^{k}_j\|,
\end{equation}
where $\mathcal{N}(i)$ represents the set of Neighbors of node $i$ (excluding $i$ itself).
Therefore, the overall loss value $loss_{semantic}$ for learning semantic graphs is defined as: 
\begin{equation}
loss_{semantic}=L_{dis}+\lambda L_{\text{node}},
\end{equation}
where $\lambda$ is a regularization factor, which controls the importance of the $L_{\text{node}}$ term compared to the $L_{dis}$ in the overall loss function.
Our disentanglement loss function, $L_{dis}$, ensures that the learned semantic graphs remain maximally separated. In practice, when using multiple graphs, we observe that one of the learned graphs serves as a reference, ensuring sufficient separation between the others. This effect arises naturally from the cosine similarity constraint, which forces one graph to act as an anchor in feature space.
Therefore, by considering one graph as an anchor, we determine the optimal number of semantic graphs, $K$, by calculating the mean Moran's I across all semantic graphs for a given $K$. This mean Moran's I serves as a summary statistic to compare different configurations of $K$. A higher mean Moran's I indicates better spatial coherence across all semantic graphs and greater biological relevance for each semantic graph.
Additionally, we incorporate another metric—the overall loss value, $loss_{semantic}$—to assist in selecting the optimal number of semantic graphs. We validated this approach on the DLPFC dataset and then applied it to all other datasets. Fig.~S10 and Fig.~S11 illustrate the Moran's I values corresponding to each semantic graph for $K=2$ to $K=7$, as well as the values of $loss_{semantic}$ for different choices of $K$, respectively.
Empirical observations indicate that $K=4$ not only provides good separation but also effectively captures the disentanglement of biological features.

\subsubsection{GNN-based encoder-decoder for learning ST embeddings}
For each generated semantic graph, we use an encoder to derive the latent embedding of the input ST data based on the corresponding graph. Specifically, we apply an encoder with GCN layers to handle $k$ semantic graphs, where each graph offers a unique set of edge weights but shares the same edge indices. Each encoder processes its assigned semantic graph to perform weighted feature aggregation.
For each semantic graph, we employ $k$ pairs of encoders, with each encoder $e$ consisting of $l$ GCN layers designed to process information from its corresponding semantic graph. The operation within each encoder's GCN layer is defined as:
\begin{equation}
Z^{(l, k)}_{e} = \sigma\left(\mathbf{d}^{k} \cdot \tilde{A} \cdot Z^{(l-1, k)}_s \cdot W^{(l-1,k)}_{e} + b^{(l-1,k)}_{e}\right),
\end{equation}
where $Z^{(l, k)}_{e}$ represents the feature vector at layer $l$ ($Z^{(0,k)}_e$ is set to normalized gene expression matrix) in the $k^{th}$ encoder, $\mathbf{d}^{k}$ denotes the edge weight corresponding to the $k^{th}$ graph descriptor, $W^{(l-1,k)}_{e}$ and $b^{(l-1,k)}_{e}$ are the trainable weight matrix and bias vector for layer $l$ in the $k^{th}$ encoder, and $\sigma(\cdot)$ denotes a non-linear activation function, such as the ReLU function.
\subsubsection{Fusion of the latent representations for each semantic graph}

Let \( Z^{(k)}_e \) represent the latent embeddings for the \( k^{\text{th}} \) semantic graph, where \( k = 1, 2, \dots, K \), and the corresponding attention weights are \( \alpha_k \). The fused embedding, \( Z_{\text{fused}} \), is computed as a weighted sum of the individual embeddings:
\begin{equation}
Z_{\text{fused}} = \sum_{k=1}^{K} \alpha_k Z^{(k)},
\end{equation}
where the attention weights \( \alpha_k \) are normalized using the softmax function across all embeddings, ensuring that:
\begin{equation}
\alpha_k = \frac{\exp(\text{attn\_score}_k)}{\sum_{k=1}^{K} \exp(\text{attn\_score}_k)}.
\end{equation}
Here, \( \text{attn\_score}_k \) represents the attention score for the \( k^{\text{th}} \) graph's embedding, which is computed by the attention mechanism. This score indicates the importance of each semantic graph in the embedding space. By using the attention mechanism, we not only adaptively aggregate the most relevant information from each semantic graph but also introduce an additional layer of interoperability to our learned semantic graph.\\
After obtaining the fussed latent representations $Z_{\text{fused}}$, we use the decoder $e$ to reconstruct the node feature matrix. The decoder's objective is to reverse the encoding process by taking the fused embedding as input and reconstructing the original features. Similar to the encoder, each layer $t$ in the decoder updates the feature representation, progressively reconstructing the original node features $X$. The operation at each layer \(t\) can be expressed as follows:
\begin{equation}
H^{t}_e = \sigma\left(\tilde{A} \cdot H^{t-1}_e \cdot W^{t-1}_ {d}+ b^{t-1}_{d}\right),
\end{equation}
where $H^{t}_e$ denotes the reconstructed gene expression matrix at the $t$-th layer, and $H^{0}_e$ is initialized as the fused embedding $Z_{\text{fused}}$. $W_{d}$ and $b_{d}$ are the trainable weight matrix and bias vector, respectively.
\subsection{SemanticST's objective function}
The objective of SemanticST consists of two loss functions. First, we aim to leverage gene expression data to incorporate biological information during training by minimizing the reconstruction loss $\ell _rec$. This loss function updates SemanticST's parameters by minimizing the distance between the reconstructed gene expression matrix $H_e$ and the normalized gene expression matrix $\tilde{X}$. The loss function can be obtained as follows:
\begin{equation}
\ell_rec=\sum_{i=1}^{N_{spot/cell}} \Vert\tilde{x}_i - h_i \Vert_2.
\end{equation}

\subsubsection{Community-based detection loss function}
Although reconstruction loss can produce meaningful embeddings and improve the informativeness of the embedding space, the unsupervised nature of the learning process in SemanticST makes it challenging to effectively learn a robust representation. To address this and learn the global structure of the data without increasing training time or introducing corrupted graphs (which occur in contrastive learning), we apply the \textit{mincut} loss function followed by the DMC learning algorithm \cite{DUONG2023109126}. This approach enhances the embedding space by incorporating spatial dependencies through the adjacency matrix, enabling the model to preserve global structure while learning discriminative features. \\
If we consider $M$ communities in the embedding space, we can assign each node in the spatial graph to the membership matrix $P\in {0,1}^{ N_{spot/cell} \times M}$, which represents the soft assignment of each node to a community. The membership matrix $P$ is obtained by applying the Gumbel-Softmax \cite{gu2018neural} to the node embeddings. To capture the affinity between different communities in the spatial graph and to compute the relationships between communities based on how strongly nodes in those communities are connected, we calculate the community association matrix $\mathbf{C}=P^T\tilde{A}P$, where the off-diagonal elements $\mathbf{C}_{ij}$ represent the inter-community association and the diagonal elements $\mathbf{C}_{ii}$ represent the internal association of a community. Thus, if $q_m=\sum^M_{j=1}\mathbf{C}_{mj}$ represents the total association of community $m$ with other communities, and $d_m=\mathbf{C}_{mm}$ represents the self-association for each community, we can minimize the mincut loss function $\ell_{mincut}$ to encourage separation between communities by pushing inter-community associations apart and pulling intra-community associations together, as follows:
\begin{equation}
\ell_{mincut}=\sum^M_{m=1}\frac{q_m-d_m}{g_m}.
\end{equation}
Since our task is unsupervised and we do not know how many clusters there are, we decided to set the number of communities to correspond with the dimensions of the hidden space. This method allows the model to optimize the division of the latent space into separate groups directly. Consequently, it enhances the interpretability and significance of the learned representations, effectively constraining the embedding space to align with these naturally defined communities.
Finally, the overall loss function of SemanticST can be defined as follows:
\begin{equation}
\ell_{overall}=\beta_1\ell_{rec}+\beta_2\ell_{mincut},
\end{equation}
where $\beta_1$ and $\beta_2$ represent the impacts of the reconstruction loss and the mincut loss, respectively. In practice, we set $\beta_1=10$ and $\beta_2=0.1$. The Adam optimizer \cite{diederik2014adam} was used as the optimization algorithm, with a learning rate of $0.001$ and a training duration of $1000$ epochs. 
\subsection{Spatial domain clustering}
After training the model, we consider the output of the decoder, $H^{t}_e$, as our final representation of the data and apply a clustering algorithm to this output. This final representation offers another perspective on the input ST data through semantic learning and serves as a higher-level representation by incorporating spatial information. For ST data where ground truth is available and the number of clusters is known, we used the mclust \cite{fraley2012mclust} algorithm. For datasets without this information, we apply the Leiden clustering algorithm from SCANPY \cite{wolf2018scanpy}. Each obtained cluster is considered a spatial domain, denoting a specific region or niche that carries similar biological information.
\subsection{Trajectory Inference}
For spatial trajectory inference, we utilize the PAGA algorithm \cite{wolf2019paga} from the SCANPY package, which is applied to the spatial domains identified through clustering on the latent representations generated by the SemanticST algorithm.
\subsection{Differentially expressed genes}
For identifying differentially expressed genes, we use the Wilcoxon test from the SCANPY package, applying an adjusted p-value threshold of $\leq 0.01$ and a log fold change threshold of $<1$. 
\subsection{Vertical and horizontal integration}
All the previously mentioned explanations focus on a single tissue slice as the input and aim to learn its latent representation. However, due to the technical limitations of ST technologies, consecutive slices are often available, both horizontally and vertically. In vertical integration tasks, the presence of tissue thickness results in vertically consecutive slices from the same tissue, necessitating a method to learn joint representations across these slices. Importantly, addressing batch effects and learning joint embeddings while mitigating these effects is crucial. To address these challenges, we extend our model to handle integration tasks both vertically and horizontally.

For vertical integration, we first align the shared genes across slices to ensure consistency and construct a uniform gene expression matrix. Inspired by SpaGCN \cite{hu2021spagcn}, we calculate the adjacency matrix for each ST dataset in the same manner as for a single slice, as described earlier. We then combine the individual adjacency matrices into a single block diagonal matrix. The resulting block diagonal matrix represents the integrated graph, where each dataset is treated as a separate component within the graph. This structure preserves the spatial relationships within each dataset while enabling integration into a unified representation. Finally, the uniform gene expression matrix and combined adjacency matrix are passed into SemanticST to learn latent representations and effectively remove batch effects. For DLPFC slices, we apply this method, while for vertical integration in breast cancer, we utilize the integrated dataset provided by GraphST. Additionally, for visualizing vertical integration, we use PASTE \cite{zeira2022alignment} in pairwise slice integration mode.
For horizontal integration, assuming two slices as an example, we align the spatial coordinates of two slices to ensure continuity and proper spatial overlap. The spatial coordinates of the first slice remain unchanged. For the second slice, the x-coordinates are shifted to match the minimum x-coordinate of the first slice, while the y-coordinates are adjusted to align with the maximum y-coordinate of the first slice. Following this alignment, we concatenated the two slices into a single integrated slice, preserving only shared gene expression data and constructing a graph adjacency matrix for the SemanticST algorithm. This integration emphasizes learning consistent representations both within and between spots or cells across slices.
\subsection{Mini-batch training}
Graph machine learning has advanced the analysis of complex graph structures, the scalability challenges---such as the neighbour explosion phenomenon \cite{hamilton2017inductive}---limit its applicability to large-scale datasets. This issue is particularly relevant in ST analysis, where modern technologies can map the locations of nearly a million cells across a tissue. Many ST analysis methods leverage spatial information to construct graph neighbourhoods, fully utilizing spatial data. However, these approaches often process the entire graph structure within the model, making them impractical for large-scale ST data such as Xenium.
Mini-batch training provides a solution but introduces additional challenges, as ST graph structures are defined by Euclidean distances, making it difficult to preserve both local and global information during training. To address this, SemanticST incorporates a mini-batch training framework, ensuring scalability for datasets with millions of samples. The dataset is divided into manageable mini-batches using PyTorch's DataLoader, with the batch size dynamically adjusted based on the dataset size. The entire dataset is used for both training and testing, with shuffling enabled during training to introduce randomness and disabled during testing to maintain consistency. SemanticST was executed on a server with Intel Xeon 56-core CPUs and NVIDIA RTX 6000 Ada Generation GPUs. For datasets with fewer than 40,000 samples, we recommend disabling mini-batch training, as they can be efficiently processed in a single batch without excessive training time. However, users can enable or disable mini-batch training based on their specific needs and system capabilities.

\section*{Data availability}
All data used in this study were obtained from previously published sources. A comprehensive list of these sources is provided in Supplementary Table S1. For convenience, we have also made the compiled data available via Zenodo: \url{https://doi.org/10.5281/zenodo.15339700}.
\section*{Code availability}
The SemanticST algorithm is an open-source Python implementation available at \url{https://github.com/roxana9/SemanticST}.
\bibliography{ref}

\end{document}